%% file: physalis_scalar.tex
\documentclass[preprint,12pt]{elsarticle}

\usepackage{graphicx}
\usepackage{amssymb}
\usepackage{amsthm}
\usepackage{amsmath}
\usepackage{psfrag}
\usepackage{color}
\usepackage{rotating}
\usepackage{pstricks}

\newcommand{\be}{\begin{equation}}
  \newcommand{\ee}{\end{equation}}
\newcommand{\eg}{e.g.}

\newcommand{\cf}{cf.~}
\newcommand{\eq}[1]{equation~(\ref{#1})}

\newcommand{\eqs}[3]{equations~(\ref{#1}) {#2}~(\ref{#3})}

\newcommand{\sect}[1]{section~\ref{#1}}

\newcommand{\fig}[1]{figure~\ref{#1}}
\newcommand{\Fig}[1]{Figure~\ref{#1}}

\journal{}
\begin{document}

\begin{frontmatter}

  % --------------------------------------------------------------
  \title{Improved procedure for the computation of Lamb's coefficients in
    the {\sc physalis} method for particle simulation}
  \author[a]{K.~Gudmundsson\corref{cor1}}
  \cortext[cor1]{Corresponding author}
  \ead{k.gudmundsson@utwente.nl}
  \address[a]{Faculty of Science and Technology and J. M. Burgers
    Centre for Fluid Dynamics, University of Twente, P.O. Box 217, 7500
    AE Enschede, The Netherlands}
  \author[a,b]{A.~Prosperetti}
  \address[b]{Department of Mechanical Engineering, The John Hopkins
    University Baltimore, MD 21218, USA}

  % --------------------------------------------------------------
  \begin{abstract}

    The {\sc physalis} method is suitable for the simulation of flows with
    suspended spherical particles. It differs from standard immersed
    boundary methods due to the use of a local spectral representation of
    the solution in the neighborhood of each particle, which is used to
    bridge the gap between the particle surface and the underlying fixed
    Cartesian grid. This analytic solution involves coefficients which are
    determined by matching with the finite-difference solution farther
    away from the particle. In the original implementation of the method
    this step was executed by solving an over-determined linear system via
    the singular-value decomposition. Here a more efficient method to
    achieve the same end is described. The basic idea is to use scalar
    products of the finite-difference solutions with spherical harmonic
    functions taken over a spherical surface concentric with the
    particle. The new approach is tested on a number of examples and is
    found to posses a comparable accuracy to the original one, but to be
    significantly faster and to require less memory. An unusual test case
    that we describe demonstrates the accuracy with which the method
    conserves the fluid angular momentum in the case of a rotating
    particle.

  \end{abstract}


\end{frontmatter}

% --------------------------------------------------------------
\section{Introduction} \label{intro}

Sediment transport, settling suspensions, fluidized beds and porous
media are but a few examples of fluid flows with fixed or moving
particles. As these examples show, such flows are frequently
encountered and of great importance in many problems of scientific and
technological relevance. From a computational perspective, they pose a
major challenge as the particles constitute a very complex and
frequently time dependent internal boundary which needs to be adequately
resolved for a {\em bona fide} simulation.

Until very recently the only practical approach was the
Eulerian-Lagrangian point-particle model, in which the particles are
assimilated to mathematical points contributing extra forces to the
fluid, while their motion is found by integrating in time a dynamic
equation in which the fluid forces are modelled rather than deduced
from first principles. The last decade has seen the development of
methods capable of dealing with the finite extent of the
particles. The most successful such method so far is the one developed
by Uhlmann \cite{Uhlmann:2005gf, Uhlmann:2008il}, which has been also
used by Lucci et al.  \cite{Lucci:2010vd}.
% and further improved
% \cite{KempeFroehlich}: An improved immersed boundary method with direct 
% forcing for the simulation of particle laden flows
% This method is of the immersed boundary type and ***

An alternative approach, dubbed {\sc physalis}, first described
in~\cite{Prosperetti:2001fj, Takagi:2003hk} and, more fully,
in~\cite{Zhang:2005ty}, differs from standard immersed boundary methods
in its reliance on a local spectral representation of the solution in
the neighborhood of each particle; a brief description is 
provided in section~\ref{framework}. Advantages of this method are the
exact satisfaction of the no-slip condition at the particle surface,
the great simplification of the calculation of the hydrodynamic forces
and couples, and the avoidance of the complex issues arising from the
lack of geometrical conformity between the curved particle boundary
and the underlying fixed Cartesian grid. Furthermore, the use of a
local spectral representation of the solution permits one to describe the
effect of each particle with fewer degrees of freedom
than conventional finite-difference-based methods. As a consequence,
the grid resolution can be kept relatively low without compromising
the accuracy of the solution.

{\sc physalis} has been thoroughly validated in the original papers
and it has proven its value in a number of applications: the
sedimentation of 1,024 particles~\cite{Zhang:2006wk}, the turbulent
flow around a fixed particle~\cite{Naso:2010ti}, the flow induced by a
rotating particle~\cite{Liu:2010tu} and over a porous
medium~\cite{Zhang:2009ky, LIU:2011hn}. An interesting variant which
uses artificial compressibility to generate a time-stepping
pressure equation has been described and used by Perrin and
Hu~\cite{Perrin:2006cl, Perrin:2008dh}.

The basic idea of {\sc physalis} is to use an analytic local solution
of the modified (Navier-\nolinebreak) Stokes equation as a bridge between the
particle boundary and the finite-difference solution. The local
analytical solution contains undetermined coefficients which are found
iteratively by matching the finite-difference solution with the local
one.

The significant improvement described in this paper concerns the
calculation of the expansion coefficients, which was carried out using
the singular-value decomposition in the original implementation. We
show how this step can be executed by taking suitable scalar products
with a significant saving in execution time. The purpose of this paper
is to describe the implementation of this new method and its
verification.

% --------------------------------------------------------------
\section{General framework} \label{framework}

We briefly review here the {\sc physalis} method for particle
simulation, as originally introduced for cylindrical particles
\cite{Prosperetti:2001fj,Takagi:2003hk}, and later extended to
spherical particles \cite{Zhang:2005ty}. 

The objective is to simulate, on a fixed Cartesian grid,
incompressible fluid flow around spherical particles, while obeying
the no-slip and no-penetration velocity boundary-conditions on the
particle surface.  To this end it is necessary to solve the
incompressible Navier-Stokes equations,
\begin{eqnarray} \label{NS}
  \rho \left[\frac{ \partial \mathbf{U}}{\partial t} + \mathbf{U}\cdot \boldsymbol{\nabla}
    \mathbf{U}\right] &=& -\boldsymbol{\nabla} p + \mu \boldsymbol{\nabla}^2 \mathbf{U} +
  \rho \mathbf{g} 
  \,,\\
  \boldsymbol{\nabla} \cdot \mathbf{U} &=& 0,
\end{eqnarray}
where $\mathbf{U}$, $\rho$, $p$, and $\mathbf{g}$ respectively denote
the velocity, density, pressure, and gravity fields. 

As with other immersed-boundary methods, since the particles do not
conform to the computational grid, the boundary conditions imposed by
the particles must be transferred to the adjacent nodes. The {\sc
  physalis} method enables this transferral to be done in a spectrally
accurate manner.

To proceed, we transform the above equations to a frame in which the
particle is at rest. (In the case of more than one particle this step
is carried for each particle.)  The fluid velocity $\mathbf{u}$ in the
particle rest-frame is related to $\mathbf{U}$, the velocity in the
laboratory frame, by
\begin{equation}
  \mathbf{U} = \mathbf{u} + \mathbf{w} + \mathbf{\Omega} \times \mathbf{x},
\end{equation}
where $\mathbf{w}$ and $\mathbf{\Omega}$ respectively represent the
linear and angular velocities of the particle, and $\mathbf{x}$ is the
position with respect to the particle center. The momentum equation in
the particle rest frame is given by
\begin{eqnarray} 
  \rho \left[ \frac{\partial \mathbf{u}}{\partial t} + \mathbf{u}
    \cdot \boldsymbol{\nabla} \mathbf{u} + 2 \,\mathbf{\Omega} \times
    \mathbf{u}\right] = -\boldsymbol{\nabla} p + \mu \boldsymbol{\nabla}^2
  \mathbf{u} + \rho \mathbf{g} -\rho \left[\mathbf{\dot{w}}
    +\mathbf{\dot{\Omega}}\times \mathbf{x} + \mathbf{\Omega} \times
    (\mathbf{\Omega} \times \mathbf{x})\right],
\end{eqnarray}
and that of continuity is given by
\begin{equation}
  \boldsymbol{\nabla} \cdot \mathbf{u} = 0.
\end{equation}
Now let 
\begin{equation}
  \mathbf{u}^\Omega = \frac{r^5-a^5}{10\nu r^3}
  \mathbf{\dot{\Omega}}\times \mathbf{x} \;\;\; \mbox{ and
  } \;\;\;p^\Omega = \frac{1}{2} \rho (\mathbf{\Omega}\times \mathbf{x})^2 -
  \rho (\mathbf{\dot{w}}-\mathbf{g})\cdot \mathbf{x},
\end{equation}
where $r = |\mathbf{x}|$, and $a$ represents the particle radius. Then,
if we introduce the modified velocity and pressure fields
\begin{equation}
  \tilde{\mathbf{u}} = \mathbf{u}-\mathbf{u}^\Omega \;\;\; \mbox{ and
  } \;\;\;  \tilde{p} =p-p^\Omega,
\end{equation} 
we can write the rest-frame momentum and continuity
equations respectively as
\begin{eqnarray}
  \rho \left[ \frac{\partial \mathbf{u}}{\partial t} + \mathbf{u} \cdot
    \boldsymbol{\nabla} \mathbf{u} + 2 \mathbf{\Omega} \times \mathbf{u}\right] &=&
  -\boldsymbol{\nabla} \tilde{p} + \mu \boldsymbol{\nabla}^2 \tilde{\mathbf{u}}, \mbox{ and } \\
  \boldsymbol{\nabla} \cdot \tilde{\mathbf{u}} &=& 0.
\end{eqnarray}
Since the term within the square brackets vanishes at the
particle surface, it will be small in its immediate
vicinity. Therefore, in this region, $\tilde{p}$ and
$\tilde{\mathbf{u}}$ will approximately satisfy the Stokes equations,
given by 
\begin{equation} \label{stokes}
  -\boldsymbol{\nabla} \tilde{p}
  + \mu \boldsymbol{\nabla}^2 \tilde{\mathbf{u}} = 0 \;\;\; \mbox{ and
  } \;\;\;\boldsymbol{\nabla} \cdot \tilde{\mathbf{u}} = 0,
\end{equation}
for which Lamb~\cite{Lamb:1932wz} developed an exact general solution for 
flow past a sphere. It should be stressed that the method is intended 
for application to finite-Reynolds-number flows. 
The procedure may be seen as a linearization of the 
Navier-Stokes equations about the solid-body motion which the fluid in 
contact with the particle surface executes at any Reynolds number 
by virtue of the no-slip condition.

The flow field is expanded in terms of vector spherical harmonics (allied 
to the standard scalar spherical harmonics), which 
form a complete basis for square-integrable vector functions on the 
sphere~\cite{Prosperetti:2011tu}. The velocity 
$\tilde{\mathbf{u}}$ is given by \cite{Lamb:1932wz, Kim:2005wv}
\begin{equation} \label{u}
  \tilde{\mathbf{u}} = \frac{\nu}{a^2}\sum_{n=-\infty}^\infty
  \frac{1}{(n+1)(2n+3)}\left[\frac{1}{2} (n+3)r^2\boldsymbol{\nabla} p_n -
    n\mathbf{x}p_n\right] + \frac{\nu}{a}\sum_{n=-\infty}^\infty \left[ a\boldsymbol{\nabla} \phi_n +
    \boldsymbol{\nabla} \times (\mathbf{x} \chi_n)\right],
\end{equation}
and the pressure $\tilde{p}$ by
\begin{equation}\label{p}
  \tilde{p} = \frac{\mu \nu}{a^2}\sum_{n=-\infty}^\infty p_n,
\end{equation}
where the $p_n$, $\phi_n$, and $\chi_n$ are solid harmonics of
order $n$:
\begin{equation} \label{solidharm}
  \left[ \begin{array}{c}
      p_n \\ \phi_n \\ \chi_n \end{array} \right]
  = \left(\frac{r}{a}\right)^n \sum_{m=-n}^n \left[ \begin{array}{c} 
      p_{nm}\\ \phi_{nm}\\ \chi_{nm} \end{array}\right]Y_n^m(\theta,\varphi)\, .
\end{equation}
Here $p_{nm},\,\phi_{nm}\,,\chi_{nm}$ are undetermined dimensionless
coefficients and the functions $Y_n^m$ are normalized spherical
harmonics which are shown in detail in the Appendix. The regular
harmonics, corresponding to non-negative $n$, represent a general
incident flow, while the singular harmonics (negative $n$) represent
the disturbance field due to the particle. Although the coefficients $\{p_{nm},
\phi_{nm},\chi_{nm}\}$ are in general complex, the relations 
$\{\overline{p}_{nm}, \overline{\phi}_{nm},\overline{\chi}_{nm}\} =
(-1)^m\{p_{n,-m}, \phi_{n,-m},\chi_{n,-m}\}$ (in which the overline 
denotes the complex conjugate) obviate the need for dealing with 
complex numbers and the entire calculation can be carried out in the real 
domain. 

The central observation at the root of the {\sc physalis} method lies in 
the recognition that, in the immediate neighborhood of the particle, the 
velocity and pressure fields must be describable as in (\ref{u}) and 
(\ref{p}). Thus, provided the coefficients are known,  (\ref{u}) and 
(\ref{p}) can be used to ``transfer'' the boundary conditions from the 
particle surface to a  \emph{cage} of adjacent grid nodes enclosing the 
particle. After this step, 
the computation is carried out only on the nodes of the fixed Cartesian 
grid and the complexity deriving from the geometric mis-match between the 
particle boundary and the grid is eliminated. 

With this background, we can now describe the algorithm:
\begin{enumerate}
\item By matching a provisional finite-difference solution (\eg, that at the 
  previous time step) to the explicit representations (\ref{u}), 
  (\ref{p}) and related ones (see below) at the cage 
  nodes, find provisional values of the expansion coefficients;
\item Use these provisional values to generate new velocity boundary 
  conditions at the cage nodes;
\item Solve again the Navier-Stokes equations on the finite-difference grid 
  using these boundary conditions;
\item Repeat to convergence.
\end{enumerate}

It is important to note that the coefficients appearing in 
(\ref{solidharm}) embody the physics of the interaction of the particle 
with the fluid and contain therefore important information. For example, 
the hydrodynamic force, $\mathbf{F}$, and couple, $\mathbf{L}$, acting 
on the particle can be expressed in terms of the $n=1$ coefficients 
as~\cite{Zhang:2005ty}: 
\begin{eqnarray}
  \mathbf{F} &=& \rho v(\mathbf{\dot{w}}-\mathbf{g}) + 6\pi \mu a
  \mathbf{\Phi} + \pi \mu a^3 \mathbf{P} \label{Force}\\
  \mathbf{L} &=& 8v\mu \mathbf{X} + \rho a^2 V \mathbf{\dot{\Omega}},\label{Couple}
\end{eqnarray}
where $v=\frac{4}{3}\pi a^3$ is the volume of the particle, and we have defined
$\mathbf{\Phi} = (\phi_{11}^r,\phi_{11}^i,\phi_{10}^r)$, $\mathbf{P} =
(p_{11}^r,p_{11}^i,p_{10}^r)$, and $\mathbf{X} =
(\chi_{11}^r,\chi_{11}^i,\chi_{10}^r)$, with superscripts $r$ and $i$
denoting the real and imaginary parts. Thus, unlike other immersed boundary 
methods, once the coefficients are known the force and couple are known 
as well with no need for additional calculations. 

As discussed in the following section, the specific contribution of
the present paper concerns the first step, that of deducing from a
given velocity and pressure field the values of the coefficients.

% --------------------------------------------------------------
\section{Computing the expansion coefficients} \label{projection} 

Away from the particle(s), the Navier-Stokes equations (\ref{NS}) are 
solved on a staggered grid by a standard finite-difference method as 
described in~\cite{Zhang:2005ty}. This reference also explains in detail 
the manner in which the \emph{cage} of Cartesian nodes enclosing the particle 
is constructed. Let ${\mathbf x}_K$, with $K=1,\,2,\, \ldots,\,N$ be 
the cage nodes. In the original algorithm, the coefficients are calculated 
by requiring that the pressure field given by (\ref{p}) and the vorticity 
field $\tilde{\boldsymbol{\omega}}$ obtained from (\ref{u}) reproduce the 
finite-difference results. In other 
words, one forms a linear system involving the coefficients as unknowns, 
e.g. 
\begin{equation} 
  \tilde{p}\,=\,  \frac{\mu\nu}{a^2}\sum_{n=-\infty}^\infty 
  \left(\frac{r_K}{a}\right)^n \sum_{m=-n}^n p_{nm}
  Y_n^m(\theta_K,\varphi_K)
  \,=\, p({\mathbf x}_K) -p^\Omega({\mathbf x}_K) \, , 
  \label{oldphys}
\end{equation}
in which $\mathbf{x}_K=(r_K,\theta_K,\phi_K)$, the left-hand side is
as given in (\ref{p}) and contains the unknown coefficients, while the
right-hand side is the finite-difference result at the cage nodes.
The vorticity field is treated similarly. As explained before, the
coefficients thus obtained are then substituted into (\ref{u}) to
generate velocity boundary conditions at the cage nodes.

In this way, four scalar equations are obtained for each cell
comprising the cage. Taken together over the cage, this set of
equations forms an overdetermined linear system for the unknown
coefficients which is solved by means of the singular value
decomposition (SVD) algorithm, i.e., in essence, in a least-squares
manner.  Although the method is stable, the typically large systems
that result (\eg, a matrix of $2400\times 46$ for a grid-resolution of
8 cells per radius, and with the analytical expressions truncated at
$N_c = 3$) prompted us to search for a more efficient procedure. 

There is, however, an alternate and perhaps more natural way to obtain
the coefficients which relies on the orthogonality property of the 
spherical harmonics, for example 
\begin{equation} \label{ortho}
  (Y_l^k,Y_n^m) \equiv \int_0^\pi \sin \theta d\theta \int_0^{2\pi}
  \overline{Y}_l^k Y_n^m  d\varphi = \delta_{ln}\delta_{km}\, .
\end{equation}
In the spirit of this 
approach, (\ref{oldphys}) is replaced by 
\be 
\label{scprd1}
\left(Y_n^m,\tilde{p}\right) \,=\,\left(Y_n^m, p-p^\Omega\right) \, .
\ee
or, upon substituting (\ref{p}) for $\tilde{p}$ and using (\ref{ortho}) 
and recalling that $Y_{-n-1}^m\,=\,Y_n^m$, 
\be
\left(Y_n^m,\tilde{p}\right)\,=\, \frac{\mu \nu}{a^2} \left(p_{nm}
  +p_{-n-1,m}\right).
\ee
It is shown in the Appendix that the coefficients of negative order can be 
expressed in terms of those of positive order to find 
\begin{eqnarray}\label{pPro}
  \frac{a^2}{\mu\nu} (Y_n^m,\tilde{p})  = 
  \left[ 1-\frac{n}{2}
    \frac{2n-1}{n+1}\left(\frac{a}{r}\right)^{2n+1}\right] 
  \left(\frac{r}{a}\right)^n p_{nm} -
  \frac{n(2n-1)(2n+1)}{n+1} \left(\frac{a}{r}\right)^{n+1} \phi_{nm}.
\end{eqnarray}
This property of the spherical harmonics to single out individual 
coefficients is analogous to the Fourier transform and, 
as discussed in \sect{implementation}, there exists a ``Fast Spherical
Transform'' (FST) \cite{Healy:2003wn}, incorporating and scaling as
the Fast Fourier Transform. As will be shown below, this
procedure proves considerably faster than the SVD approach of
\cite{Zhang:2005ty}, and comparably accurate. 

The other coefficients can be determined by a similar procedure. However 
there are different possible choices of field variables to use for this 
purpose, which we now discuss. 

In the derivations we will use two orthogonality properties of the vector 
spherical harmonics similar to (\ref{ortho}), namely (see 
e.g. Ref.~\cite{Prosperetti:2011tu})
\begin{equation}\label{rgradortho}
  (r\boldsymbol{\nabla} Y_l^k, r \boldsymbol{\nabla} Y_n^m) =
  r^2  \int_0^\pi \sin \theta d\theta \int_0^{2\pi}
  \boldsymbol{\nabla} \overline{Y}_l^k \cdot\boldsymbol{\nabla}
  Y_n^m  d\varphi = n(n+1)\delta_{ln}\delta_{km},
\end{equation}
\begin{equation}\label{gradortho}
  (\mathbf{x} \times \boldsymbol{\nabla} Y_l^k,\mathbf{x} \times 
  \boldsymbol{\nabla} Y_n^m) =
  \int_0^\pi \sin \theta d\theta \int_0^{2\pi}
  (\mathbf{x} \times \boldsymbol{\nabla} \overline{Y}_l^k)\cdot 
  (\mathbf{x} \times \boldsymbol{\nabla}
  Y_n^m)  d\varphi = n(n+1)\delta_{ln}\delta_{km}.
\end{equation}

It is found that, whatever field one may consider,  
$\phi_{nm}$ and $p_{nm}$ appear coupled, while $\chi_{nm}$ appears 
separately. 

% --------------------------------------------------------------
\subsection{Computing $\phi_{nm}$ and $p_{nm}$}\label{phiANDp}

In general it is desirable to choose pressure as one of the field variables
used to calculate the coefficients $p_{nm}$ and $\phi_{nm}$, since 
this field is smoother than the velocity field and, additionally, does 
not vanish near the particle. Thus, (\ref{pPro}) is one of the equations 
that we use to determine the coefficients. 

A combination of any two out of three other scalar products, involving
the radial component of velocity $\tilde{u}_r =
\mathbf{e}_r \cdot \tilde{\mathbf{u}}$; the transversal components of
velocity $\tilde{\mathbf{u}}_\perp=
\tilde{\mathbf{u}}-\tilde{u}_r\mathbf{e}_r$; or the transversal
components of vorticity $\tilde{\boldsymbol{\omega}}_\perp =
\tilde{\boldsymbol{\omega}} -\tilde{\omega}_r\mathbf{e}_r$ may be used to 
solve for $\phi_{nm}$ and $p_{nm}$. We will present each of these options 
below, and, in \sect{results}, demonstrate their respective performance.

\begin{enumerate}

\item {\it Radial component of velocity:} we take the scalar 
  product of $Y_n^m$ and $ \tilde{u}_r$ as found from (\ref{u}), obtaining
  \begin{eqnarray}\label{urPro} 
    \frac{a}{\nu}(Y_n^m,\tilde{u}_r) &=&
    \frac{n}{4}\left[\frac{2}{2n+3}+\left(\frac{2n+1}{2n+3}\frac{a^2}{r^2}-1
      \right) \left(\frac{a}{r} \right)^{2n+1} \right] 
    \left(\frac{r}{a}\right)^{n+1} p_{nm} 
    \nonumber\\ && \quad+n\left[ 1+\frac{1}{2}\left[
        2n-1-(2n+1)\frac{r^2}{a^2}\right]
      \left(\frac{a}{r}\right)^{2n+1}\right]  
    \left(\frac{r}{a}\right)^{n-1} \phi_{nm}, 
  \end{eqnarray} 
  where we have used (\ref{ortho}) and the relations (\ref{prel}) and 
  (\ref{phirel}) to eliminate the coefficients with negative index. 

\item {\it Transversal components of velocity:}  By using (\ref{rgradortho}) 
  in (\ref{u}) we find 
  \begin{eqnarray}\label{uperpPro} 
    \frac{a}{\nu}(r\boldsymbol{\nabla} Y_n^m,\tilde{\mathbf{u}}_\perp) &=&
    \frac{n}{4}
    \left\{ \frac{2(n+3)}{2n+3} + n\left[(n-2)-\frac{2n+1}{2n+3}
        \frac{a^2}{r^2} \right]\left(\frac{a}{r}\right)^{2n+1}\right\}
    \left(\frac{r}{a}\right)^{n+1} p_{nm}  + \nonumber \\ && \quad
    n\left[n+1+\frac{1}{2}\left[ (n-2)(2n+1)\frac{r^2}{a^2}-n(2n-1)\right]
      \left( \frac{a}{r}\right)^{2n+1} \right]  
    \left(\frac{r}{a}\right)^{n-1}\phi_{nm}
  \end{eqnarray} 
  where again we have used the relations (\ref{prel}) and 
  (\ref{phirel}) to eliminate the coefficients with negative index. 

\item {\it Transversal components of vorticity:} Upon taking the curl of
  the velocity given by (\ref{u}) and projecting on the transversal 
  direction we find 
  \begin{eqnarray} \label{omegaPerp}
    \frac{a^2}{\nu}  \tilde{\boldsymbol{\omega}}_\perp
    &=& \mathbf{e}_r \times \sum_{n=1}^\infty \frac{1}{n+1}\left\{ \left[
        1+\frac{2n-1}{2}\left(\frac{a}{r}\right)^{2n+1}\right]
      \boldsymbol{\nabla}^\prime p_n + (2n-1)(2n+1)
      \left(\frac{a}{r}\right)^{2n+1}\boldsymbol{\nabla}^\prime \phi_n \right\}
    \nonumber \\ && \qquad \qquad 
    + \frac{a}{r}\sum_{n=1}^\infty \left[
      n+1+n\left(\frac{a}{r}\right)^{2n+1}\right]
    \boldsymbol{\nabla}^\prime \chi_n,\label{transvort}
  \end{eqnarray}
  where for brevity we have defined the operator $\boldsymbol{\nabla}^\prime$ as,
  \eg, 
  \begin{eqnarray} 
    \boldsymbol{\nabla}^\prime p_n = \left(\frac{r}{a}\right)^n r 
    \sum_{m=-n}^n p_{nm} \boldsymbol{\nabla}  Y_n^m(\theta,\varphi). 
  \end{eqnarray}
  Upon taking the scalar product of this equation with 
  $\mathbf{x} \times \boldsymbol{\nabla} Y_l^k$ and using (\ref{gradortho}) 
  we obtain 
  \begin{equation}\label{wPro}
    \frac{a^2}{\nu}  (\mathbf{x} \times \boldsymbol{\nabla} Y_n^m, 
    \tilde{\boldsymbol{\omega}}_\perp) =  n
    \left[ 1+\frac{2n-1}{2}\left(\frac{a}{r}\right)^{2n+1} \right]
    \left(\frac{a}{r}\right)^n p_{nm} + n(2n-1)(2n+1) 
    \left(\frac{a}{r}\right)^{n+1} \phi_{nm}, 
  \end{equation}
  where again (\ref{prel}) and (\ref{phirel}) have been used to eliminate the 
  coefficients with negative index.  

\end{enumerate}
Having in hand \eq{pPro}, along with either (\ref{urPro}),
(\ref{uperpPro}), or (\ref{wPro}), we can now solve for the two
unknowns, $\phi_{nm}$ and $p_{nm}$.

% --------------------------------------------------------------
\subsection{Computing $\chi_{nm}$}\label{ComputingChi}
Three scalar products can be formed that involve $\chi_{nm}$:

\begin{enumerate}

\item {\it Transversal components of velocity:} projecting
  $\tilde{\mathbf{u}}_\perp$ upon $\mathbf{x} \times
  \boldsymbol{\nabla} Y_n^m$, rather than $r\boldsymbol{\nabla}
  Y_n^m$, as done in \eq{uperpPro}, we obtain
  \begin{equation} \label{uperpProChi}
    \frac{a}{\nu}(\mathbf{x} \times \boldsymbol{\nabla} Y_n^m,\tilde{\mathbf{u}}_\perp)
    = n(n+1)\left[1- \left(\frac{a}{r}\right)^{2n+1}\right] 
    \left(\frac{r}{a}\right)^n
    \chi_{nm},
  \end{equation}
  where we have used \eq{chirel}. 

\item {\it Radial component of vorticity:} $\tilde{\omega_r}$ is given by
  \begin{equation}
    \frac{a^2}{\nu}\tilde{\omega}_r = \frac{a}{r} \sum_{n=1}^\infty
    n(n+1)\left[1-\left(\frac{a}{r}\right)^{2n+1}\right]\chi_n.
  \end{equation}
  Projecting this expression upon $Y_n^m$, we obtain
  \begin{equation} \label{omegarPro}
    \frac{a^2}{\nu}(Y_n^m,\tilde{\omega}_r) = n(n+1) \left[1-
      \left(\frac{a}{r}\right)^{2n+1}\right] 
    \left(\frac{r}{a}\right)^{n-1}  \chi_{nm},
  \end{equation}

\item {\it Transversal components of vorticity:} projecting
  $\tilde{\boldsymbol{\omega}}_\perp$ upon $r\boldsymbol{\nabla} Y_n^m$,
  rather than $\mathbf{x} \times \boldsymbol{\nabla} Y_n^m$, as was
  the case in \eq{wPro}, we obtain
  \begin{equation} \label{chiPro}
    \frac{a^2}{\nu} (r\boldsymbol{\nabla} Y_n^m,\tilde{\boldsymbol{\omega}}_\perp) = n(n+1)
    \left[n+1+n\left(\frac{a}{r}\right)^{2n+1}\right] 
    \left(\frac{r}{a}\right)^{n-1} \chi_{nm}.
  \end{equation}
\end{enumerate}
Any one of relations (\ref{uperpProChi}), (\ref{omegarPro}), and
(\ref{chiPro}),  may be used to obtain  $\chi_{nm}$.

% --------------------------------------------------------------
\section{Implementation}\label{implementation}

The second-order accurate discretization of the Navier-Stokes
equations, along with the construction of the cage and the manner in
which the particle boundary conditions are applied at the cage
surface, are described in detail in \cite{Zhang:2005ty}. We will
therefore focus our discussion here on those aspects of the present
methodology that diverge from the original version, namely, the
calculation of the expansion coefficients.

The scalar products appearing in (\ref{scprd1}) and the following
equations amount to integrations carried out over a spherical surface
concentric with and enclosing the particle. In general, therefore, the
quadrature nodes do not coincide with the finite-difference grid
points. We thus require a procedure for approximating the values of
$\boldsymbol{\omega}$, $\mathbf{u}$, and $p$ at the quadrature nodes
on the basis of those on the finite-difference grid.  To this end we
adopt two separate approaches: interpolation, and local
linearization. For the reasons explained below, we limit the latter to
those scalar products involving $\tilde{p}$ and
$\tilde{\mathbf{u}}_\perp$ in relations (\ref{pPro}),
(\ref{uperpPro}), and (\ref{uperpProChi}).

% --------------------------------------------------------------
\subsection{Interpolation}
Since the spatial and temporal discretization is second-order
accurate, an interpolation of the same order or higher is
appropriate. We experimented with both cubic splines and quadratic
Lagrange-polynomials, both implemented in the GNU Scientific Library
\cite{Anonymous:wz}. These proved comparably accurate, but the
Lagrange interpolation is considerably more expedient and was hence
adopted.

Given the point $\hat{\mathbf{x}} = (\hat{x},\hat{y},\hat{z})$ at which the 
value of e.g. the pressure $p$ is required, the interpolation proceeds as 
follows:
\begin{enumerate}
\item Let $\mathbf{x}_{i,j,k}$ be the cell containing $\hat{\mathbf{x}}$;
\item Centered on, and including this cell, construct a cube of 27
  cells having coordinates $\mathbf{x}_{i+\alpha,j+\beta,k+\gamma}$,
  with $\alpha,\,\beta,\,\gamma=0,\,\pm1$, at  which $p$ is known; 
\item Arbitrarily choosing to interpolate first on the $y$-$z$ planes
  of the cube, we  interpolate $p(x_i,y_j,z_k)$ in $z$ for each $y_j$ and
  $x_i$, and evaluate the results at $z = \hat{z}$, yielding
  $p(x_i,y_j,\hat{z})$;
\item Interpolate $p(x_i,y_j,\hat{z})$ in $y$ for each $x_j$,
  and evaluate at $y = \hat{y}$, yielding $p(x_i,\hat{y},\hat{z})$;
\item A final interpolation in $x$ is performed and evaluated at $x=
  \hat{x}$, resulting in the desired approximation of
  $p(\hat{\mathbf{x}})$.
\end{enumerate}
This three-dimensional, second-order accurate interpolation therefore requires 
13 one-dimensional interpolations in total. 

% --------------------------------------------------------------
\subsection{Local linearization} \label{seloclin}
For pressure, we write 
\begin{equation} \label{TaylorP}
  p(\hat{\mathbf{x}}) \approx p(\mathbf{x}_c) +
  \left. \boldsymbol{\nabla} p \right|_{\mathbf{x}_c}
  \cdot (\hat{\mathbf{x}}-\mathbf{x}_c),
\end{equation}
where $\hat{\mathbf{x}}$ is defined as above, and 
$\mathbf{x}_c=\mathbf{x}_{i,j,k}$ is
the cell-center which, given the staggered-grid formulation, is where
$p$ resides. The gradient $\boldsymbol{\nabla} p$ is
approximated with second-order accurate (at $\mathbf{x}_c$) central
differences, using adjoining cells. 

For velocity, the corresponding expression is 
\begin{equation} \label{TaylorU}
  \mathbf{u}(\hat{\mathbf{x}}) \approx
  \mathbf{u}(\mathbf{x}_c) + {\sf J}(\mathbf{x}_c) 
  (\hat{\mathbf{x}}-\mathbf{x}_c)  \;\;\; \mbox{ where } \;\;\;
  {\sf J}_{ij}(\mathbf{x}_c) =
  \left. \frac{\partial u_i}{\partial x_j} \right|_{\mathbf{x}_c}.
\end{equation}
Since the velocity is defined at face centers, we approximate
$\mathbf{u}(\mathbf{x}_c)$, to second-order accuracy, with the
average across the cell. The gradients ${\sf J}_{ij}$ are calculated with
central differences. The diagonal components ${\sf J}_{ii}(\mathbf{x}_c)$
reside at the cell-center. As the remaining components reside at
edge-centers, we approximate them to second-order accuracy with the
average across the cell.

While the quadratic interpolation has higher accuracy than the linearization, 
the latter has some advantages that make it worth exploring. Firstly, note 
that the estimate $\mathbf{u}(\hat{\mathbf{x}}) $ is divergence-free 
to the same accuracy as the velocity field at the nodes, since
\begin{equation} \label{divfree}
  \boldsymbol{\nabla}\cdot \mathbf{u}(\hat{\mathbf{x}}) =
  {\sf J}_{ii}(\mathbf{x}_c) =
  \left. \frac{\partial u_i}{\partial x_i}
  \right|_{\mathbf{x}_c},
\end{equation}
while the same is not necessarily true of the interpolation.
Secondly, the operation count involved with the linearization is lower
than for the interpolation. Lastly, the spatial extent of the region
involved with the computation is reduced, as only the six adjacent, 
face-sharing cells need be considered.

The same approach applied to the vorticity would only give a zero-order
accuracy. An effort to improve this low accuracy, on the other hand, 
would require a comparatively much larger set of cells. This is the 
reason why we did not pursue this method for the vorticity options. 

With these procedures for sampling the flow field at the quadrature points 
on the sphere, the next task is to evaluate the scalar
products. The present integrands (\eg, in \eq{ortho}) are oscillatory
and increasingly so with higher $n$. Particular care must therefore be
exercised in distributing the quadrature points in order to maintain
stability. Given the homogeneity of the azimuthal ($\phi$) direction,
a uniform grid $\phi_k = 2\pi k/2N$, is appropriate, where $k =
0,\hdots,2N-1$ and $2N$ is the number of quadrature points in
$\phi$. The natural distribution in the polar ($\theta$) direction,
spanned by the associated Legendre polynomials (\cf
\eqs{Ydef}{and}{Pdef}), is $\theta_j = \pi (2j+1)/4N$, where $j$ has the 
same range as $k$.  

The scalar product on the sphere can essentially be considered as the
product of discrete Fourier and Legendre transforms in the azimuthal
and polar directions, respectively. Using the quadrature points
described above, the cost of a direct numerical evaluation of these
transforms scales as $O(N^4)$. Faster alternatives exist, however. The
Fast Fourier Transform can naturally be used in the azimuthal
direction, and there additionally exists a Fast Legendre Transform,
due to \cite{Healy:2003wn}. The combination of these fast algorithms
reduces the cost scaling to $O(N^2 \log(N))$. The authors of
\cite{Healy:2003wn} have released a C-implementation of their method,
the {\it SpharmonicKit}\footnote{See
  http://www.cs.dartmouth.edu/{\raise.17ex\hbox{$\scriptstyle\sim$}}geelong/sphere/},
which we embed in our code.

Lastly, we note that the Fast Spherical Transform of \cite{Healy:2003wn} 
includes only the 
spherical harmonics $Y_n^m$, and not gradients thereof. To compute the
scalar products involving $\boldsymbol{\nabla}Y_n^m$ (\eg,
\eq{uperpProChi}), we express these in terms of the $Y_n^m$:
\begin{eqnarray}
  \frac{\partial Y_n^m}{\partial \phi} &=& im Y_n^m,\\
  \frac{\partial Y_n^m}{\partial \theta} &=& \frac{m}{\tan{(\theta)}}
  Y_n^m + \frac{N_{n,m}}{N_{n,m+1}} e^{-i\phi} Y_n^{m+1}.
\end{eqnarray}
It should be noted that, in spite of the explicit appearance of the imaginary
unit, the actual computations are carried out in the real domain only.

% --------------------------------------------------------------
\section{Results} \label{results}

In this section we contrast the various aspects of the singular-value
decomposition (SVD) and the scalar product (SP) version of {\sc
  physalis}. In particular we examine relative speed and accuracy, and
address the question of which SP is best suited.

% --------------------------------------------------------------
\subsection{Comparing the speed of the SVD and SP}\label{speed}

As mentioned in \sect{projection}, \cite{Zhang:2005ty} compute the
expansion coefficients by matching the vorticity and pressure, based
respectively on the analytical expansions in \eqs{u}{and}{p}, to those
from the finite-difference field, resulting in a linear system, ${\sf A}
\mathbf{c} = \mathbf{b}$. ${\sf A}$ is a rectangular matrix of size $L
\times M$, where $L = N_\omega+N_p$ and $M = 3N_c(N_c+2)+1$, 
where $n = N_c$ is the degree at which the summations in \eqs{u}{and}{p} are
truncated; $\mathbf{c}$ is the vector containing the unknown expansion
coefficients; and $\mathbf{b}$ contains the values from the
finite-difference field.  $N_\omega$ and $N_p$ respectively represent
the number of vorticity and pressure nodes comprising the cage.

The commonly used SVD algorithm can be shown \cite{Lawson:1974wg} 
to scale asymptotically as $O(LM^2+M^3)$. In contrast, the cost of the
FST scales as $O(N^2 \log(N))$, where $2N\times2N$ is the number of
samples.  If the flow variables $\tilde{\boldsymbol{\omega}}$,
$\tilde{\mathbf{u}}$, and $\tilde{p}$ are {\it band-limited} so that
$n \le B$, then $N = B$ will suffice to calculate
the coefficients without error (this is analogous to the
Nyquist-Shannon sampling theorem \cite{Shannon:1949fb}), assuming that
the samples are error-free  \cite{Healy:2003wn}. 
In the general case we do not know $B$
and, therefore, to prevent aliasing, take $N = 3 N_c$. 
In terms of $N_c$, the cost of the FST therefore scales
similarly.

\begin{figure} 
  \begin{center} 
    \input{CostScaling_Nc.tex}
    \includegraphics{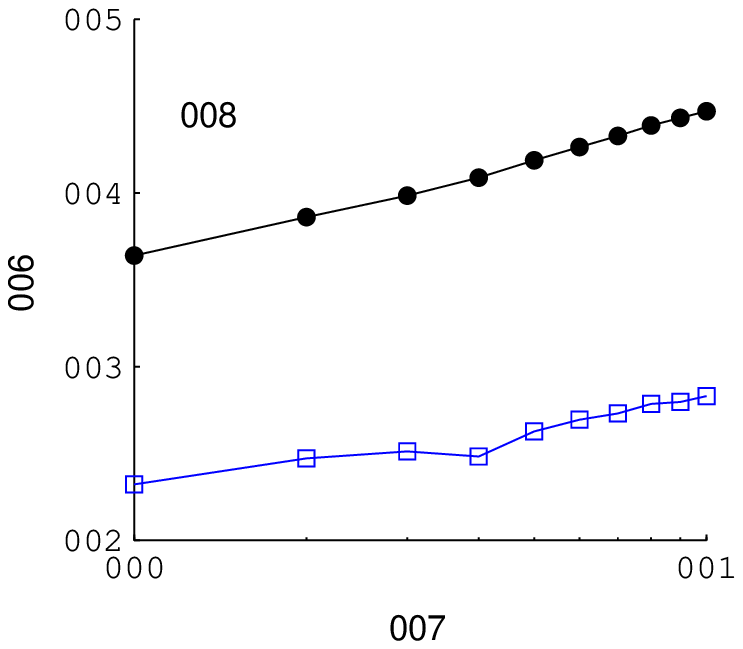}
    \input{CostScaling_ah.tex}
    \includegraphics{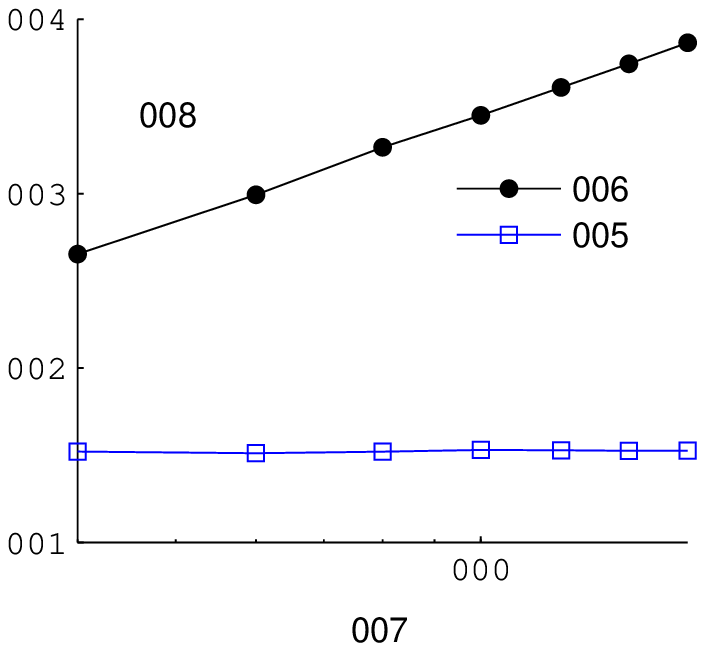}
    \caption{Cost of coefficient recovery using the SVD and SP methods,
      as a function of $N_c$ (left), and grid resolution $a/h$
      (right). The timings were measured empirically using an Intel
      2.66GHz CPU.}
    \label{CostScaling} 
  \end{center}
\end{figure}

As shown in \cite{Zhang:2005ty}, $N_c = 4$ suffices to adequately
describe flows with particle Reynolds numbers up to about 100, with a
grid resolution of $a/h = 8$, where $h$ is the grid spacing. As such,
the asymptotic scalings above are not as relevant as empirical
measurements of the cost associated with a single calculation of the
coefficients. Such measurements are shown in \fig{CostScaling}, both
as a function of $N_c$ (with fixed $a/h$) and of the grid resolution
$a/h$ (with fixed $N_c$). The SP approach proves over two orders of
magnitude faster than the SVD at all $N_c$ shown. We also note that
the cost of both the interpolation and the linearization, described in
the previous section, is independent of the grid resolution as the 
number of quadrature points is only determined by $N_c$. This is
apparent from \fig{CostScaling}, right, while the cost of the SVD
grows strongly with $a/h$.

While the SP has been shown to be significantly faster than the SVD,
there are some caveats of note. Firstly, the most expensive portion of
the overall calculation is the solution of the Poisson equation for
pressure. The speed-up in the calculation of the coefficients, while
impressive, will therefore only affect the overall calculation
fractionally, and not proportionally. Notwithstanding, the overall
gains can be significant, particularly in the presence of many
particles. Secondly, we note that the SP is only faster for moving
particles. Since the cage of stationary particles is fixed, the SVD
needs only to be computed once, and the resulting decomposition can be
re-used, resulting in comparable speed with the SP. If the particle is
moving, on the other hand, the cage moves with it (once the particle
moves a distance of $h/2$ from the center of the cage, a new cage is
constructed \cite{Zhang:2005ty}), requiring a call to the
SVD. Although this does not happen at every time step, the effect can
be nonetheless significant as shown in \sect{falling}.

% --------------------------------------------------------------
\subsection{Recovery of assigned coefficients in Stokes
  flow} \label{recovery}

The simplest test of the accuracy with which the coefficients are
calculated can be carried out on a flow field generated analytically
by using the expressions (\ref{u}) and (\ref{p}). We use these
relations with arbitrarily chosen coefficients to assign the fields at
the grid points. We choose $N_c = 4$ (for a total of $3N_c(N_c+2)+1 =
73$ coefficients assigned), and a spatial resolution of $a/h =
10$. The SPs are computed at $r/a = 1.5$. Given that the entire field
in this case is Stokes, the solution is, in theory, insensitive to the
choice of $r/a$. However, the further out one is from the surface of
the particle, the finer the Cartesian grid becomes relative to the
scale ($r/a$) on which the $Y_n^m$ vary. Later we will demonstrate how
the accuracy varies with $r/a$ for flows in which the Stokes region
has a finite extent.

To quantify the accuracy with which the assigned coefficients are 
recovered from the assigned fields, we define an error metric as
\begin{equation} \label{rmsError}
  E_p(n) = \sqrt{ \frac{1}{n} \sum_{m=-n}^{n} |p_{nm}^A-p_{nm}|^2 },
\end{equation}
where $p_{nm}^A$ and $p_{nm}$ respectively denote the assigned and
recovered values; $E_\phi$ and $E_\chi$ are defined in the same way.

\begin{figure} 
  \begin{center} 
    \input{Recovery_p.tex}
    \includegraphics{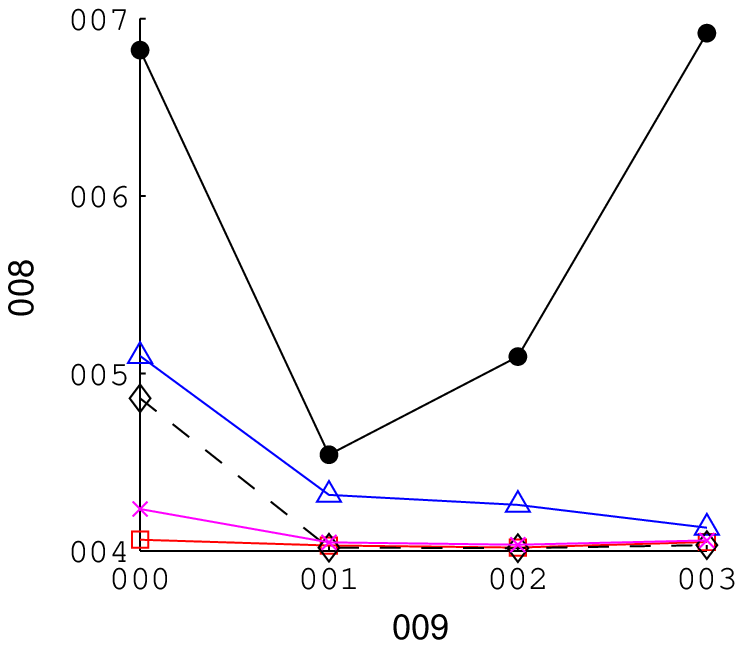}
    \input{Recovery_phi.tex}
    \includegraphics{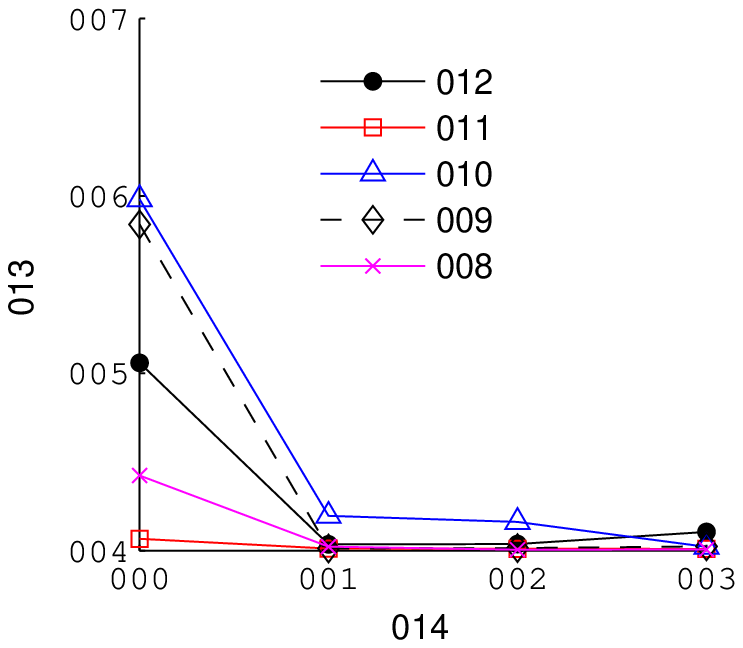}\\
    \input{Recovery_chi.tex}
    \includegraphics{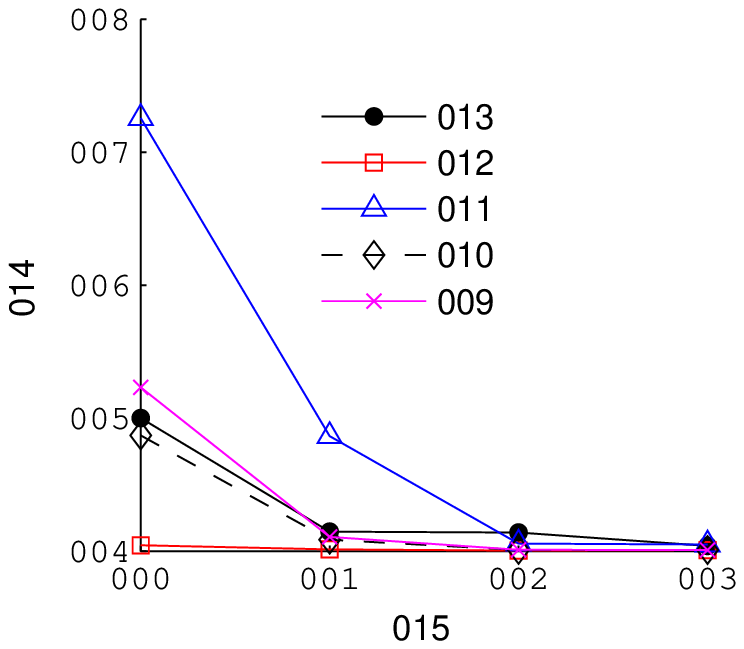}
    \caption{Here we examine the accuracy with which the coefficients
      $p_{nm}, \phi_{nm}$ and $\chi_{nm}$ are recovered from a Stokes
      flow constructed by assigning coefficients in \eqs{u}{and}{p}.
      All coefficients are set simultaneously and the error measured
      according to \eq{rmsError}. Parameters: $N_c = 4$, $a/h = 10$,
      and SPs taken at $r/a = 1.5$.}
    \label{Recovery}
  \end{center}
\end{figure}

\Fig{Recovery} shows the results so obtained. Generally, the recovery
by all methods is reasonable, with a maximum error level of $E_p =
6\%$, achieved by the SVD. In fact, all of the SPs prove more accurate
than the SVD under the $E_p$  error metric. In contrast the SVD  
incurs a larger error 
at higher values of $n$. This instability, presumably associated with the 
numerical implementation of the SVD, is 
particularly apparent in $E_p$, but can also be noted in $E_\phi$ and
$E_\chi$. In terms of $E_p$ and $E_\phi$, the SP involving the
interpolated $\tilde{\mathbf{u}}_\perp$ proves most accurate, followed
by those involving $\tilde{u}_r$ and $\tilde{\boldsymbol{\omega}}_\perp$, and
the linearization estimate of $\tilde{\mathbf{u}}_\perp$ (recall, from
\sect{implementation}, that linearization is only used with
$\tilde{\mathbf{u}}_\perp$ and $\tilde{p}$). The interpolated
$\tilde{\mathbf{u}}_\perp$ also performs best in terms of $E_\chi$,
but now followed by $\tilde{\boldsymbol{\omega}}_\perp$, $\tilde{\omega}_r$,
and the linearized $\tilde{\mathbf{u}}_\perp$. 

Although the reasons for the differing performance, in terms $E_p$,
$E_\phi$, and $E_\chi$, are somewhat unclear, we can make some
observations. Firstly, we note that the interpolated
$\tilde{\mathbf{u}}_\perp$ is likely more accurate than $\tilde{u}_r$
due to its vector nature: the relevant SPs
(\eqs{uperpPro}{and}{uperpProChi}) involve two velocity components,
presumably resulting in some cancellation of error. While the same is
true of $\tilde{\boldsymbol{\omega}}_\perp$, it should be remembered that this
is a derived quantity and some error will be associated with the
differentiation of the velocity field. The results involving the
linearized approximations in \eqs{TaylorP}{and}{TaylorU} are likely
inferior to the interpolated ones due to first-order accuracy of the
former. Stokes flow with $N_c = 4$ contains relatively rapid
variations, which are better captured by the second-order method. This
is evidenced by setting $N_c = 2$, at which degree the two methods
produce nearly the same results (not shown).  A further confirmation
is provided by the results of the next sections where the two methods
show comparable performance in smoother flows in which only the
lowest order coefficients ($n\le 2$) prove significant.

% --------------------------------------------------------------
\subsection{Drag on a sphere in a triply-periodic array} \label{gradP}

In this section we examine the accuracy of the SVD and SPs in a
pressure-driven flow through a triply-periodic array of spheres. As in
\cite{Zhang:2005ty}, we write the pressure gradient as
\begin{equation}
  \boldsymbol{\nabla}p = P\mathbf{e}_z +
  \boldsymbol{\nabla}\hat{p},
\end{equation}
in which $\hat{p}$ is periodic, $\mathbf{e}_z$ is the unit vector in
the flow direction $z$, and $P$ is a constant. By applying a momentum
balance in the flow direction at steady state, it can be shown that
the force $\mathbf{F} = F_z \mathbf{e}_z$ on the particle is given
by
\begin{equation} \label{PeriodicForce}
  F_z = (1-\beta)L^3 P, 
\end{equation}
where $\beta = \frac{4}{3}\pi a^3/L^3$. We use a periodic domain of size $L =
4a$, with the particle fixed at the center. The flow is initially
stationary, at which time a pressure gradient $P_* = a^3P/\mu\nu = 10$
is applied. This accelerates the flow until it reaches a steady state
where the drag force on the particle matches that applied by the
pressure gradient, resulting in a Reynolds number of $Re = 2aU/\nu =
22.6$, where $U$ is the mean flow rate. The accuracy of the force
calculation in \eq{Force} can then be gauged by comparing with the
exact result in \eq{PeriodicForce}. 

\begin{figure} 
  \begin{center} 
    \input{PeriodicArray_Nc.tex}
    \includegraphics{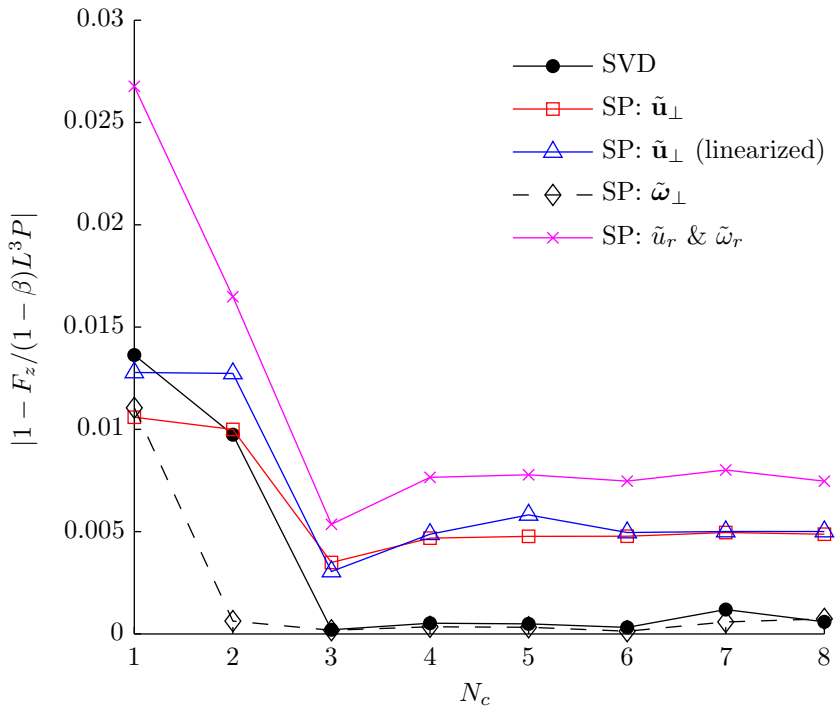}
    \caption{Here we compute a momentum balance on a single sphere in a
      triply-periodic array of spheres with an imposed pressure
      gradient, and compare with the analytical result in
      \eq{PeriodicForce}, as a function of $N_c$. Parameters: $L=4a,
      a/h = 8$, and SPs computed at $r/a = 1.25$.}  
    \label{PeriodicArray_Nc} 
  \end{center}
\end{figure}

Using a resolution of $a/h = 8$, we calculate the error
$|1-F_z/(1-\beta)L^3P|$ as a function of $N_c$, with the SPs being
computed at $r/a = 1.25$.  \Fig{PeriodicArray_Nc} show the results so
obtained. Good results are obtained for all methods, with the error
being less than $1\%$ for $N_c \ge 3$, as was the case in
\cite{Zhang:2005ty}. Notable is the performance of SP involving
$\tilde{\boldsymbol{\omega}}_\perp$, suffering the lowest error out of all the
methods, including the SVD. The two SPs involving
$\tilde{\mathbf{u}}_\perp$ now prove comparably accurate, with the
interpolated one only having a slight advantage. This is likely due to the
smoothness of the flow field, as the $n=1$ coefficients are
dominant. Incidentally, these are the coefficients involved with the
force calculation, as can be seen from \eq{Force}. In contrast, the SP
involving $\tilde{u}_r$ suffers the largest error of all. While all
velocities vanish at the surface of the particle, the radial component
does so more rapidly than the transversal ones. This causes larger
errors in the interpolation, as the signal to noise ratio is
diminished.  This explains why $\tilde{u}_r$ proves less accurate here
than in the previous section, where the SPs were computed at $r/a =
1.5$.

Since the SPs are computed at a given distance from the particle
surface, this test case provides an opportunity to explore the extent
of the Stokes region, at least for this flow
configuration. As one moves closer to the surface, SPs involving
velocity will at some point be increasingly in error as the velocity
vanishes. The SPs involving vorticity and pressure will also suffer
error in this region, as neither is continuous through the surface of
the cage where the boundary conditions are applied. Thus, upon moving
towards the cage, one will eventually interpolate through this
discontinuity, resulting in error. The error will also grow as we move
further from the particle and out of the Stokes region.

\begin{figure} 
  \begin{center} 
    \input{PeriodicArray_ra.tex}
    \includegraphics{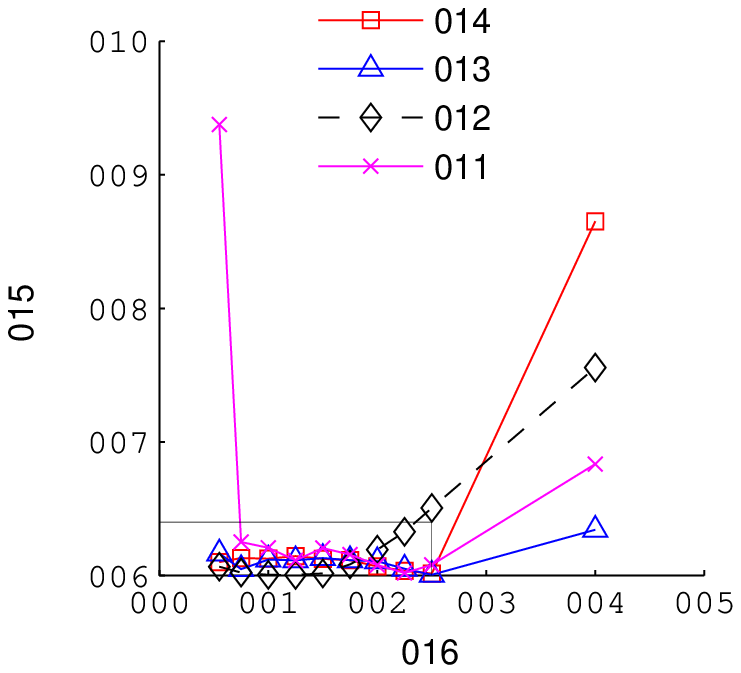}
    \input{PeriodicArray_raz.tex}
    \includegraphics{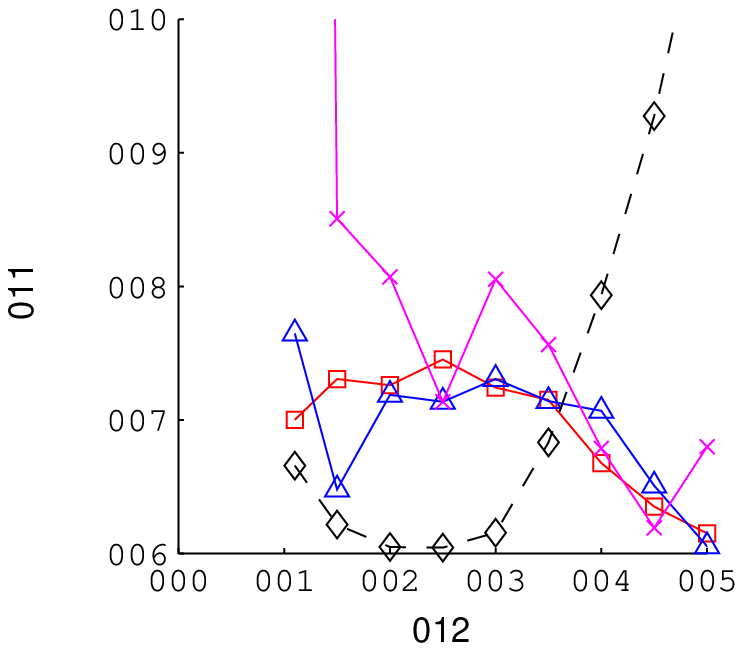}
    \caption{Error computed as in \fig{PeriodicArray_Nc}, but as a
      function of the $r/a$ at which the SPs are computed, with fixed
      $N_c=3$. The figure on the right shows a close-up view of the
      rectangular region on the left.}
    \label{PeriodicArray_ra} 
  \end{center}
\end{figure}

\Fig{PeriodicArray_ra} shows error calculations analogous to those in
\fig{PeriodicArray_Nc}, but now as a function of $r/a$, with $N_c$
fixed at 3. The behavior of the different SPs with $r/a$ is not
uniform, with $\tilde{\boldsymbol{\omega}}_\perp$ having a minimum in error at
$r/a = 1.25$, while the minimum for those involving velocities occurs 
close to $r/a =1.5$. 

Lastly, we note that, as this test case is nearly axisymmetric in the 
vicinity of the particle, the coefficients $\chi_{nm}$ are very small 
and as such we cannot draw conclusions as to which SP serves best to compute 
this. This is addressed in the next subsection.

% ---------------------------------------------------
\subsection{Couple on a sphere in a triply-periodic
  array} \label{rotating}

In the previous section we presented results demonstrating the
accuracy with which the coefficient-calculation of the SVD and
different SPs conserves the momentum of flow driven by a pressure
gradient. In this section we perform an analogous set of calculations,
but this time involving the conservation of angular momentum. This 
is a stringent test of which few examples, if any, can be found in the 
literature. 

A particle rotated at a constant rate $\boldsymbol{\Omega}$ will impart
angular momentum to the surrounding fluid. In turn, in a finite or periodic 
domain, at steady state the couple at the surface of the particle 
must be balanced by the couple on the fluid on the outer boundary. 
The calculation proceeds as follows. 

We take the cross-product of $\mathbf{x}$, the position vector
relative to the particle center, with the momentum \eq{NS} at
steady state and in the absence of gravity, and integrate over the domain 
\begin{equation} \label{int}
  \rho \int_V  \mathbf{x}\times   \boldsymbol{\nabla}\cdot(  \mathbf{U} \mathbf{U})
  \,dV=
  \int_V \mathbf{x}\times(\boldsymbol{\nabla}\cdot 
  \boldsymbol{\sigma}) \,dV,
\end{equation}
where 
\begin{equation}\label{sigma}
  \sigma_{ij} = -p\delta_{ij} + \mu\left(\frac{\partial
      U_i}{\partial x_j} + \frac{\partial U_j}{\partial x_i}\right).
\end{equation}
By the symmetry of the stress tensor we have 
\begin{eqnarray}
  \left.\mathbf{x}\times(\boldsymbol{\nabla}\cdot
    \boldsymbol{\sigma})\right|_i = \epsilon_{ijk} x_j\frac{\partial
    \sigma_{kl}}{\partial x_l}
  &=& \frac{\partial}{\partial x_l} (\epsilon_{ijk}
  x_j\sigma_{kl})- \epsilon_{ijk} \sigma_{kj}
  \nonumber\\
  &=&\left.\boldsymbol{\nabla}\cdot(\mathbf{x}\times
    \boldsymbol{\sigma}\right)|_i \, .
\end{eqnarray}
An application of the divergence theorem transforms equation (\ref{int}) 
to 
\begin{equation} \label{surf}
  \rho \int_S  \mathbf{x}\times \mathbf{U} (\mathbf{U}\cdot\mathbf{n})
  \,dS=
  \int_S \mathbf{x}\times(\boldsymbol{\sigma}\cdot\mathbf{n})
  \,dS +\int_{S_p} \mathbf{x}\times(\boldsymbol{\sigma}\cdot\mathbf{n})
  \,dS,
\end{equation}
where $S$ is the surface of the domain and $S_p$ that of the particle; 
$\mathbf{n}$ is the normal vector directed out of the fluid. We have 
dropped the integral over the particle surface in the left-hand side as the 
radial component of velocity vanishes there. For periodic boundary 
conditions, which we assume, the remaining integral in the left-hand-side 
also vanishes. Furthermore
\be
\int_{S_p} \mathbf{x}\times(\boldsymbol{\sigma}\cdot\mathbf{n})
\,dS \,=\, -\mathbf{L}_p
\ee
is just the hydrodynamic couple on the particle (with the minus sign 
due to the direction of the normal) and \eq{surf} therefore becomes 
\begin{equation} \label{surf2}
  \mathbf{L}_p \,=\, \int_S \mathbf{x}\times(\boldsymbol{\sigma}\cdot\mathbf{n})
  \,dS \,\equiv \,\mathbf{L}_b \, .
  \ee

  We take  $\boldsymbol{\Omega} = \Omega_z \mathbf{e}_z$ so that 
  $\mathbf{L}_p\,=\,L_p \mathbf{e}_z$ and 
  $\mathbf{L}_b\,=\,L_b \mathbf{e}_z$. We  use 
  a domain of size $ 4a$, a grid resolution of $a/h = 8$, and a
  rotation rate of $Re = a^2\Omega_z/\nu = 20$. We compute the error
  $|1-L_b/L_p|$, as a function of $N_c$, analogously with the previous
  section. \Fig{rot_couple} shows the results. With the
  exception of $\tilde{\boldsymbol{\omega}}_\perp$, all SPs prove more accurate
  than the SVD for $N_c \ge 2$. At first glance, this seems at odds with
  the results of the previous section, where $\tilde{\boldsymbol{\omega}}_\perp$
  was the most accurate SP. However, with $\tilde{\boldsymbol{\omega}}
  = \Omega_z \mathbf{e}_z$, $\tilde{\omega}_\phi = 0$ and, therefore,
  $\tilde{\boldsymbol{\omega}}_\perp = \tilde{\omega}_\theta \mathbf{e}_\theta$. Hence, 
  the second component of this SP is lost. Further, a significant portion
  of $\tilde{\omega}_\theta$ is associated with the solid-body rotation of the
  particle, which gets subtracted out in the frame of the particle,
  decreasing the signal-to-noise ratio in the calculation of
  $\tilde{\omega}_\theta$, which in turn causes higher error levels. In
  contrast, the SP associated with $\tilde{\omega}_r$ now performs best out of
  all candidates, which is also at odds with the results of the previous
  section. As in the previous section, the SPs involving the
  interpolated, and linearized, estimates of
  $\tilde{\mathbf{u}}_\perp$ provide practically the same results.

We also compute the above error for fixed $N_c = 3$, but varying the
$r/a$ at which the SPs are computed. \Fig{rot_rOa} shows the results
so obtained. As was the case in \fig{PeriodicArray_ra}, the error for
different SPs does not behave uniformly in $r/a$. Generally, though,
it seems that there is some leeway in the placement of $r/a$, with
$1.1 \le r/a \le 1.5$ providing reasonable results.

  The difference between the performance of the various SPs in this
  section and the previous section is instructive. Clearly, which SP
  will perform best depends on the flow conditions. If, for example,
  $\tilde{\omega}_r$ is very small, it will not deliver a reliable estimate of
  the $\chi_{nm}$ (via \eq{omegarPro}). On the whole, it
  seems that the combination of $\tilde{p}$ and
  $\tilde{\mathbf{u}}_\perp$ provides consistently good estimates across
  widely varying flow conditions. One would further suggest using the
  linearization over the interpolation, given the simplicity of the
  former and its other advantages described in
  section~\ref{implementation}. 

  \begin{figure} 
    \begin{center} 
      \input{rot_couple.tex}
      \includegraphics{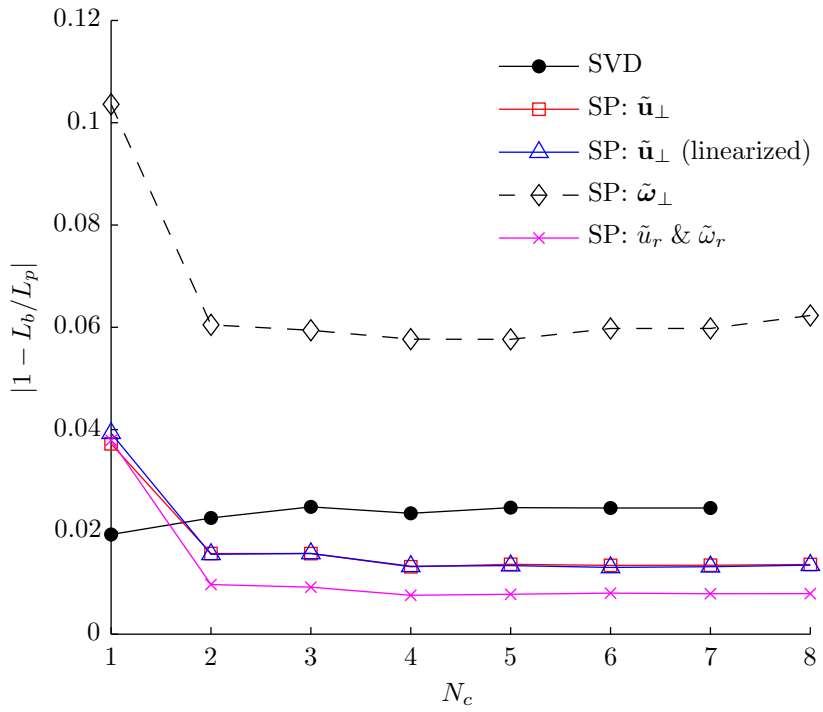}
      \caption{Here we compute the error in the calculation of couple on
        a particle rotating at constant velocity, as a function of
        $N_c$. The results for SVD at $N_c = 8$ are absent due to
        convergence failure. Parameters: $L=4a, a/h = 8$, and SPs
        computed at $r/a = 1.25$.}
      \label{rot_couple} 
    \end{center}
  \end{figure}

  \begin{figure} 
    \begin{center} 
      \input{rot_rOa.tex}
      \includegraphics{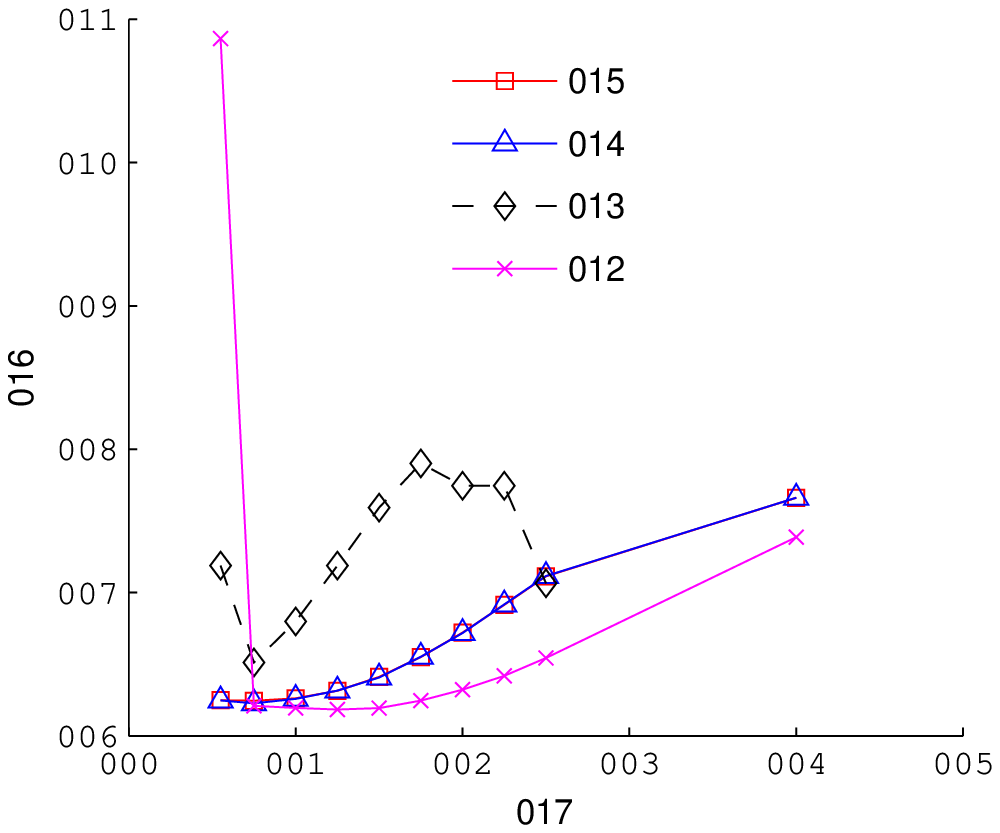}
      \caption{Error in the calculation of couple on a particle rotating
        at constant velocity, as a function of the $r/a$ at which the
        SPs are computed. The results for $\tilde{\boldsymbol{\omega}}_\perp$ at
        $r/a = 1.8$ are absent due to convergence failure. Parameters:
        $L=4a, a/h = 8$, and $N_c = 3$.}
      \label{rot_rOa} 
    \end{center}
  \end{figure}

  % --------------------------------------------------------------
  \subsection{A falling particle} \label{falling}

  The test cases in the previous sections were all for non-translating
particles. We now describe some tests on a freely falling particle in
a parallelepipedal container of length $20a$ and cross section
$8a\times 8a$.  The boundary conditions in the lateral and vertical
directions are periodic and no-slip, respectively.  The ratio of the
particle density to the fluid density is $\rho_p/\rho=2$, the
kinematic viscosity is $\nu = 0.2731$ m$^2$/s, and the acceleration of
gravity has the standard value $\mathbf{g}\cdot\mathbf{e}_z$=-9.81
m/s$^2$. The simulation is carried out with $a/h=6$ and 8,
respectively using $N_c=2$ and 3. The time-step in all cases was
$\nu\Delta t/a^2 = 0.005$, resulting in a CFL number of 0.45 at $a/h =
8$.  In light of the previous results, we narrow our scope of SPs to
those involving the linearization of $\tilde{\mathbf{u}}_\perp$ and
$\tilde{p}$, and those involving the quadratically interpolated
$\tilde{\boldsymbol{\omega}}_\perp$ and $\tilde{p}$.

  The particle is released from rest at $t = 0$. If no provision is made, at 
  $t=0$ the weight of the particle would be suddenly imposed on the fluid, 
  thus generating strong force oscillations. To avoid this artifact 
  we have ramped up the value of the acceleration of gravity in an exponential 
  manner over a time $0\leq \nu t/a^2\leq 0.1$. 

  \Fig{falling_re} shows
  the evolution of the Reynolds number $Re = 2aw/\nu$. In an unbounded
  fluid the steady state Reynolds number predicted by the Schiller-Naumann
  correlation~\cite{Schiller:1933vh} is 25, but, due to the lateral
  constraint, here it is close to 22.5 in all cases. 
  Essentially, the effect of the constraint can be viewed as a blockage effect 
  due to the image particles. While the steady-state Reynolds number varies
  slightly between the different methods and at the two different
  resolutions, the largest difference is less than 3\%.

  \begin{figure} 
    \begin{center} 
      \input{falling_re1.tex}
      \includegraphics{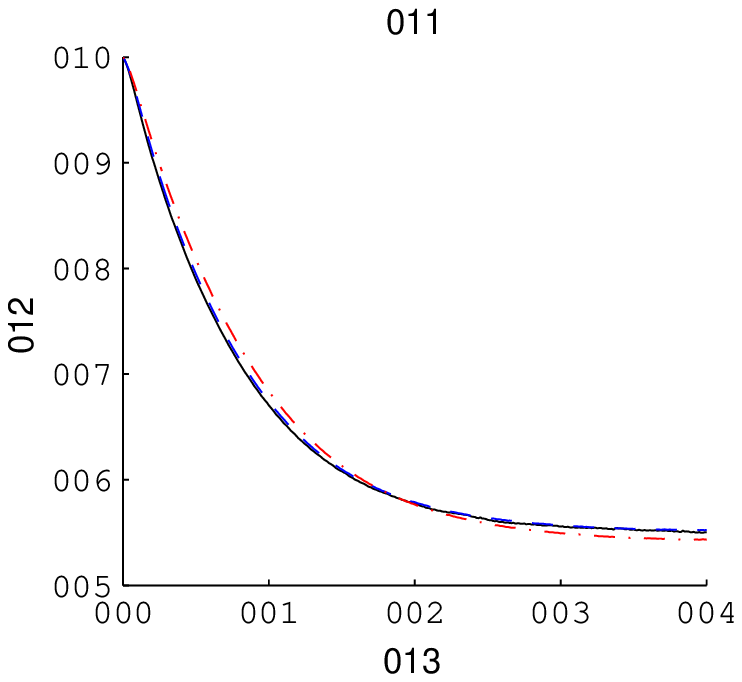}
      \input{falling_re2.tex}
      \includegraphics{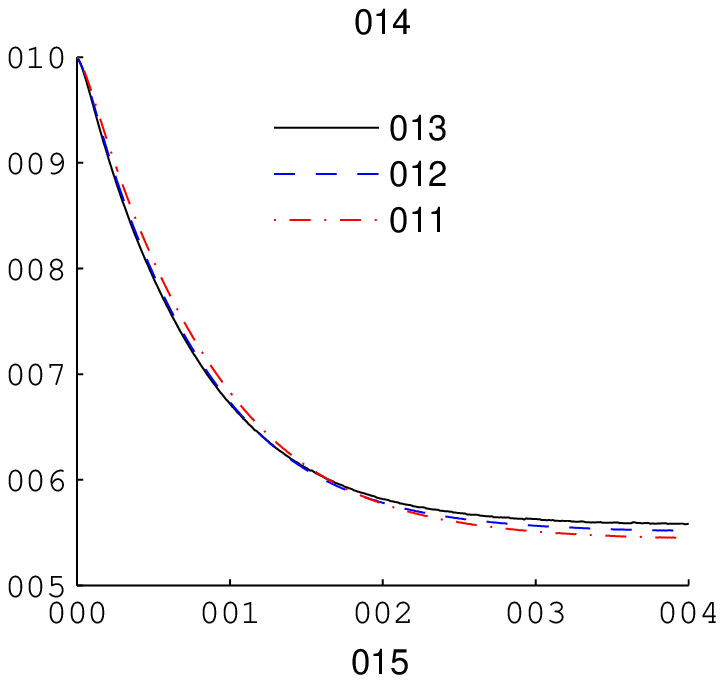}
      \caption{Evolution of the particle Reynolds number $Re = 2aw/\nu$,
        for two levels of spatial resolution.}
      \label{falling_re}
    \end{center}
  \end{figure}

  Since the cage changes with time, for the reasons explained in
  \sect{speed}, this is also a good test case to examine the CPU-time
  used by particle-related operations.  \Fig{falling_t} shows the
  cumulative time in seconds thus expended, as 
measured using an Intel 2.66GHz CPU. As expected, both SPs provide a
  significant speed-up, particularly for $a/h = 8$. For $a/h = 6$, the
  SPs have cumulatively cost roughly 14 s at $\nu t/a^2 = 1$, while the SVD costs
  roughly 109 s, larger by a  factor of 8. The cost associated with the SVD
  increases by a factor of 6 upon increasing the resolution to $a/h =
  8$, while that of the SPs only increases by a factor of 2 (recall
  that $N_c = 2$ and 3 respectively for $a/h = 6$ and 8). This is
  consistent with the time measurements demonstrated in
  \fig{CostScaling}. The cumulative savings in the cost of particle-related
  operations would be proportionally increased in the presence of
  multiple particles. Lastly, the cost associated with the SP based on the
  linearized velocity and pressure is roughly 50\% less than that based on the
  interpolated transversal vorticity and pressure.

  \begin{figure} 
    \begin{center} 
      \input{falling_t1.tex}
      \includegraphics{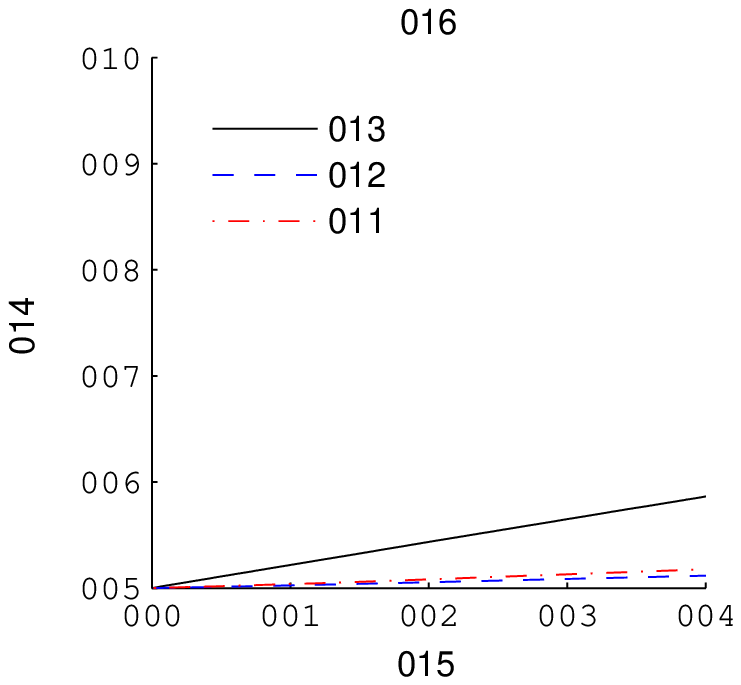}
      \input{falling_t2.tex}
      \includegraphics{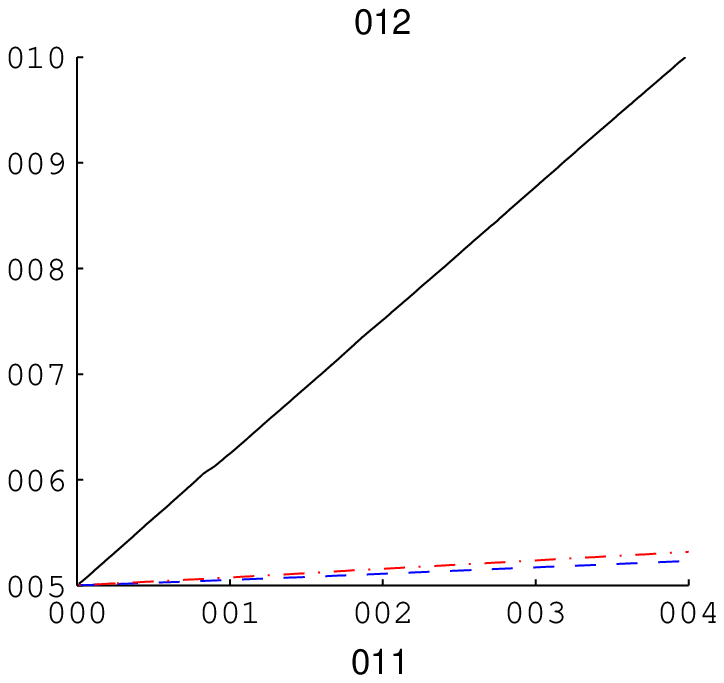}
      \caption{Cumulative time spent on particle-related operations.}
      \label{falling_t}
    \end{center}
  \end{figure}

  It is generally the case with immersed-boundary methods that force 
  oscillations are produced as the particle translates with respect to the 
  underlying grid. It was shown in \cite{Zhang:2005ty} that 
  {\sc physalis} is not immune from this problem, and it is therefore 
  of interest to examine the performance of the SPs also in this 
  respect. 

  Generally speaking, the origin of the force oscillations must be sought 
  in the minor discontinuities that the calculated force undergoes as 
  the particle moves. The same mechanism is responsible for the force 
  oscillations encountered upon a sudden release of the particle as 
  mentioned before. A discontinuity in the hydrodynamic force on the 
  particle, by the action-reaction principle, causes an impulsive force 
  on the fluid, which responds with a pressure impulse. This is the
  reason why the artificial compressibility approach of 
  \cite{Perrin:2006cl, Perrin:2008dh} suffers from this difficulty much
  less than the original {\sc physalis}. 

  In our method, at the end of each time-step, the position of the
  particle center is compared with that of the cage center and, if the
  difference is greater than $h/2$, the cage is displaced
  \cite{Zhang:2005ty}. As a consequence, a different set of nodes is
  used to compute the coefficients at the next time step, and this
  causes an unavoidable small discontinuity in the force calculation
  even though the displacement of the cage is small enough that both the
  old and the new cages are entirely in the Stokes region. This
  mechanism leads one to expect that the force oscillations decrease
  with a refinement of the grid, as it was indeed observed in
  \cite{Zhang:2005ty} and as will be shown presently. Another
  contributing factor may be the fact that the location at which the
  boundary conditions are applied on the fluid is different when the
  cage is moved. Furthermore, some of the grid points that were
  previously inside the cage become fluid points and need to be
  re-initialized as such, as described by \cite{Zhang:2005ty}. However,
  this is likely not a significant source of force oscillations due to
  the iterations in the inner loop (steps 1-4 in \sect{framework}).

  \Fig{falling_force} shows the time evolution of the
  normalized force on the particle for $a/h=6$ and 8. It should be noted
  that, for clarity, the curves for the original SVD method and
  the SP with $\tilde{\boldsymbol{\omega}}_\perp$ have been translated
  by $\pm 0.1$. The expected decrease of the amplitude of oscillations
  with increasing grid refinement is evident for all three methods,
  although the SVD curves show significant spikes even with the finer
  grid. The amplitude of the oscillations with both SP methods is
  smaller, does not present any spikes and exhibits a greater reduction
  on the finer grid. The greatest reduction is obtained by the SP with
  the linearized $\tilde{\mathbf{u}}_\perp$ which thus is seen to
  perform well also from this point of view.  

  \begin{figure} 
    \begin{center} 
      \input{falling_force1.tex}
      \includegraphics{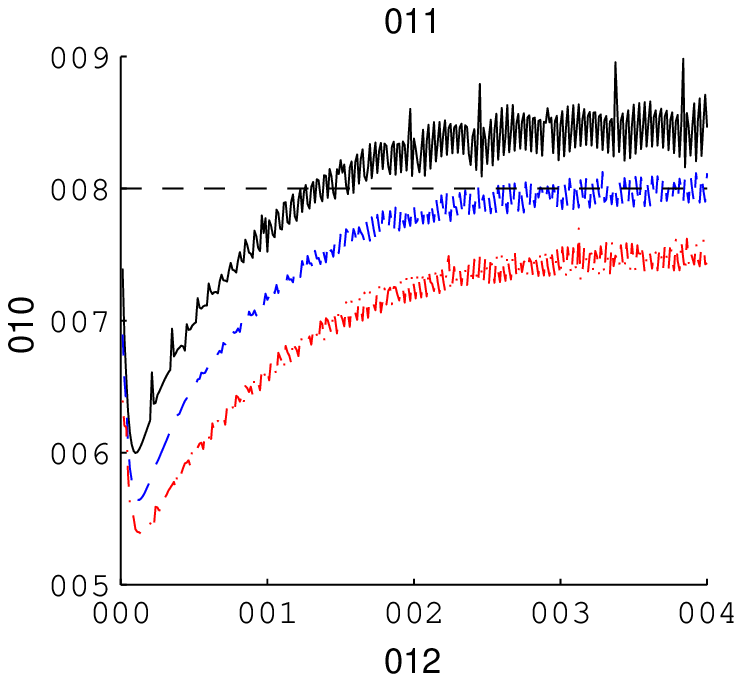}
      \input{falling_force2.tex}
      \includegraphics{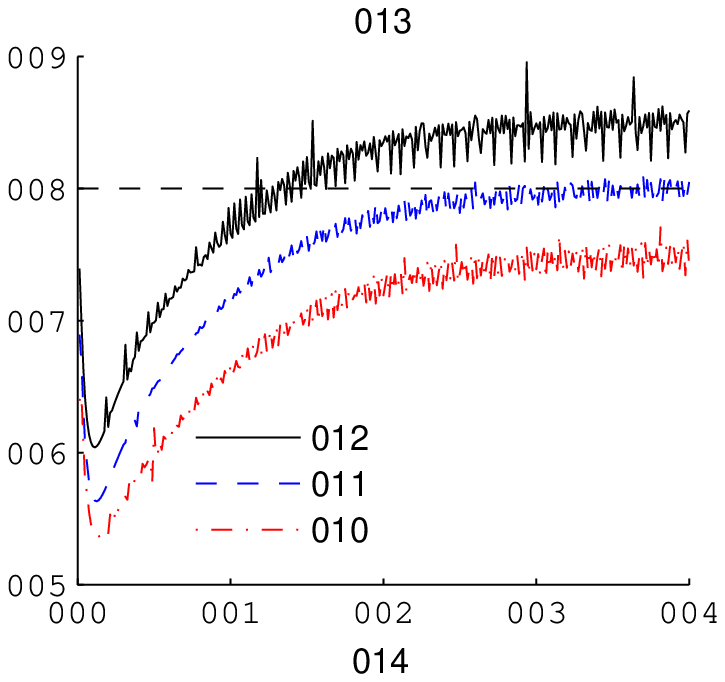}
      \caption{Evolution of normalized hydrodynamic force in the
        direction of gravity, for a particle falling from rest between
        two plates, with periodic boundary conditions in the directions
        normal to gravity. For readability, the curves representing the
        SVD and the SP involving $\tilde{\boldsymbol{\omega}}_\perp$ are
        respectively shifted by $+0.1$ and $-0.1$.}
      \label{falling_force}
    \end{center}
  \end{figure}

  % --------------------------------------------------------------
  \subsection{Taking the scalar products closer to the
    particle-surface}\label{SPcloser}

  In the above sections, the SPs were taken over a sphere with a radius 
  $r$ with $r/a>1 +ch/a$ with $1\leq c \leq 4$, where $a$ is the radius of 
  the particle. While this does not arise concerns in 
  single-particle simulations, it might under circumstances in which
  particles are expected to be in close proximity (such as for flow
  through a tightly-packed porous medium) and even to collide. In this
  section we show that $c \geq 1$ is not an intrinsic limitation of our
  method and that even $r/a = 1$  is permissible, allowing inter-particle
  contact.

  As noted before, vorticity and pressure are not continuous through the
  cage. This is the reason why, for a given cage size $R_c/a$ and grid 
  resolution, there exists a lower limit on $c$ below 
  which the calculation of vorticity and pressure becomes in
  error. While this is the case for both the interpolation and local
  linearization, the latter is less affected due the smaller
  number of cells involved, thus allowing a smaller $c$ before the cage is
  breached.

  The constraint of operating outside the cage can be met 
  either by using $r/a$ appreciably greater than 1, as done before, 
  or by decreasing $R_c/a$ as we now show. The latter option is permissible 
  since the analytic expressions (\ref{u}) and (\ref{p}) embody a smooth 
  analytic continuation of the pressure and velocity fields inside the 
  particle. Thus, the boundary 
  conditions on the cage may be applied on the inside as well. 

  As shown
  in \fig{PeriodicArray_rcage}, by taking $R_c/a = 0.9$, the SP
  involving $\tilde{\boldsymbol{\omega}}_\perp$ gives excellent results even
  with $r/a = 1$. As the velocities vanish at $r/a = 1$, the relevant
  SPs are increasingly in error close to the surface, even with $R_c/a <
  1$. In situations where $r/a < 1.05$ is required, one would thus be
  advised to use $\tilde{\boldsymbol{\omega}}_\perp$. Actually, in all the examples 
  considered, except that of section~\ref{rotating}, this option gives 
  excellent results and the associated computational overhead is only modestly
  larger than for the linearization method. 

  \begin{figure} 
    \begin{center} 
      \input{PeriodicArray_rcage.tex}
      \includegraphics{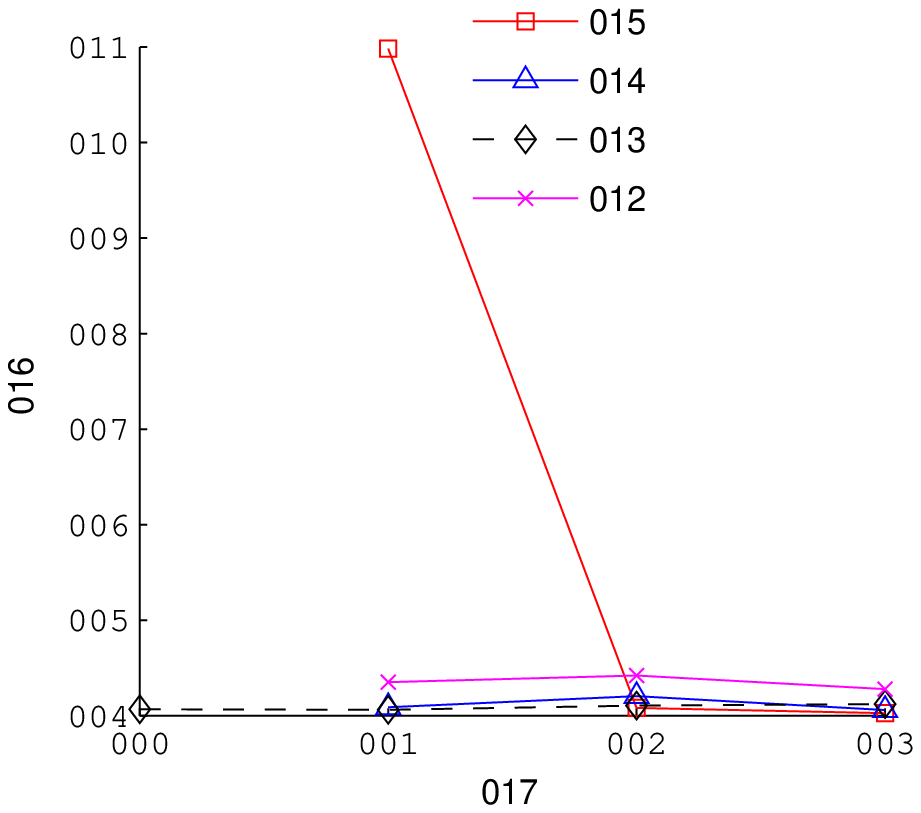}
      \caption{Error for the same flow-configuration as in \sect{gradP},
        as a function of the $r/a$ at which the SPs are taken, and the
        size of the cage, $R_c/a$. Note that the SP involving
        $\tilde{\boldsymbol{\omega}}_\perp$ can be taken right at the particle
        surface, granted that the cage is inside the particle.}
      \label{PeriodicArray_rcage}
    \end{center}
  \end{figure}

  % --------------------------------------------------------------
  \section{Summary and conclusions} \label{summary}

  In this work we have presented a new approach to calculating the
  unknown expansion coefficients in Lamb's solution for the
  near-particle flow field in the {\sc physalis} method. In the previous
  approach of \cite{Zhang:2005ty}, the coefficients were obtained via
  the SVD solution of a set of linear equations formed by equating the
  analytical expansion, evaluated at the nodes of the finite-difference
  field around the particle, to the finite-difference field itself
  (cf.~\eq{oldphys}). In the new approach we take advantage of the
  orthogonality of the set of vector harmonics by taking
  suitable scalar products thereof with values obtained from the
  finite-difference field.

  We have described how the finite-difference field can be estimated on the
  spherical surface over which the scalar products are taken and further
  how the relevant integrals are computed efficiently via a combination
  of discrete Fourier and Legendre transforms on the sphere. 

  There are several options for which fields should be used to calculate the 
  coefficients. Our results suggest that the local linearization method 
  based on the pressure and the transverse velocity 
  (section~\ref{seloclin}) is efficient and accurate provided the scalar 
  product can be computed on a surface with a radius exceeding that 
  of the particle by 1-2 mesh lengths. For simulations where the particles 
  are very close or can collide, a more accurate option would be 
  the use of the transverse vorticity in addition to the pressure
  (sections~\ref{phiANDp} and~\ref{ComputingChi}). 

  Via various examples involving stationary, rotating, and sedimenting
  particles, we have demonstrated that the new approach is more
  accurate, more stable in terms of force-oscillations, and
  significantly faster than the SVD approach of \cite{Zhang:2005ty}. The
  speed-up is particularly significant at higher Reynolds
  numbers, as $h\sim Re^{-3/4}$. As the number of points comprising the
  cage scales as $Re^{3/2}$, the cost of the SVD scales as $Re^{3/2}
  N_c^4$, (cf.~\sect{speed}). While the same resolution is of course
  required when using the scalar product, the associated cost scales
  only as $N_c^2 \log N_c$. The speed-up reported in this paper are
  therefore likely to be further enhanced at higher Reynolds numbers.

  % --------------------------------------------------------------
  \section*{Acknowledgements}
  This work was supported by the IMPACT Institute, the Netherlands.

  % --------------------------------------------------------------
  \section*{Appendix}

  The $Y_n^m$, in \eq{solidharm} and others, are the orthogonal
  spherical harmonics,
  \begin{equation}\label{Ydef}
    Y_n^m(\theta,\phi) = N_n^m P_n^m(\cos\theta) e^{i m\phi},
    \;\;\;N_n^m = \sqrt{\frac{2n+1}{4\pi}\frac{(n-m)!}{(n+m)!}},
  \end{equation}
  where $\theta$ and $\varphi$ are respectively the polar and azimuthal
  angles, and $m=0,\pm 1,\hdots,\pm n$. For $m\ge 0$, the associated Legendre
  polynomials $P_n^m$ are given by
  \begin{equation}\label{Pdef}
    P_n^m(\cos \theta) = (-1)^m (\sin{\theta})^m
    \frac{d^m}{d(\cos{\theta})^m} P_n(\cos{\theta}). 
  \end{equation}
  The constant $N_n^m$ is included in order to render the $Y_n^m$
  orthonormal. 

  The regular and singular solid harmonics can be related through the
  no-slip boundary condition on the particle surface. The resulting
  expressions %(\cf \cite{Kim:2005wv} for further details) 
  are
  \begin{eqnarray} \label{relationship}
    p_{-n-1} &=& -\frac{1}{2} \frac{2n-1}{n+1} n\left[p_n + 2
      (2n+1)\phi_n\right] \left(\frac{a}{r}\right)^{2n+1} \label{prel}\\
    \phi_{-n-1} &=& -\frac{1}{4}\frac{n}{n+1}\left[\frac{2n+1}{2n+3} p_n +
      2(2n-1)\phi_n\right]\left(\frac{a}{r}\right)^{2n+1} \label{phirel}\mbox{, and }\\
    \chi_{-n-1}& =& -\left(\frac{a}{r}\right)^{2n+1} \chi_n.\label{chirel}
  \end{eqnarray}

  \bibliographystyle{elsarticle-num}
  \bibliography{physalis_scalar.bib}

\end{document}

%% file: CostScaling_Nc.tex
% Generated using matlabfrag
% Version: v0.6.16
% Version Date: 04-Apr-2010
% Author: Zebb Prime
%
%% <text>
%
\providecommand\matlabtextA{\color[rgb]{0.000,0.000,0.000}\fontsize{10}{10}\selectfont\strut}%
\psfrag{006}[bc][bc]{\matlabtextA Cost [sec]}%
\psfrag{007}[tc][tc]{\matlabtextA $N_c$}%
\psfrag{008}[cl][cl]{\matlabtextA $a/h=6$}%
%
%% </text>
%
%% <xtick>
%
\def\matlabfragNegXTick{\mathord{\makebox[0pt][r]{$-$}}}
\psfrag{000}[ct][ct]{\matlabtextA $10^{0}$}%
\psfrag{001}[ct][ct]{\matlabtextA $10^{1}$}%
%
%% </xtick>
%
%% <ytick>
%
\psfrag{002}[rc][rc]{\matlabtextA $10^{-4}$}%
\psfrag{003}[rc][rc]{\matlabtextA $10^{-2}$}%
\psfrag{004}[rc][rc]{\matlabtextA $10^{0}$}%
\psfrag{005}[rc][rc]{\matlabtextA $10^{2}$}%
%
%% </ytick>

%% file: CostScaling_ah.tex
% Generated using matlabfrag
% Version: v0.6.16
% Version Date: 04-Apr-2010
% Author: Zebb Prime
%
%% <text>
%
\providecommand\matlabtextA{\color[rgb]{0.000,0.000,0.000}\fontsize{10}{10}\selectfont\strut}%
\psfrag{005}[cl][cl]{\matlabtextA SP}%
\psfrag{006}[cl][cl]{\matlabtextA SVD}%
\psfrag{007}[tc][tc]{\matlabtextA $a/h$}%
\psfrag{008}[cl][cl]{\matlabtextA $N_c = 3$}%
%
%% </text>
%
%% <xtick>
%
\def\matlabfragNegXTick{\mathord{\makebox[0pt][r]{$-$}}}
\psfrag{000}[ct][ct]{\matlabtextA $10^{1}$}%
%
%% </xtick>
%
%% <ytick>
%
\psfrag{001}[rc][rc]{\matlabtextA $10^{-4}$}%
\psfrag{002}[rc][rc]{\matlabtextA $10^{-2}$}%
\psfrag{003}[rc][rc]{\matlabtextA $10^{0}$}%
\psfrag{004}[rc][rc]{\matlabtextA $10^{2}$}%
%
%% </ytick>

%% file: Recovery_p.tex
% Generated using matlabfrag
% Version: v0.6.16
% Version Date: 04-Apr-2010
% Author: Zebb Prime
%
%% <text>
%
\providecommand\matlabtextA{\color[rgb]{0.000,0.000,0.000}\fontsize{10}{10}\selectfont\strut}%
\psfrag{008}[bc][bc]{\matlabtextA \begin{rotate}{-90}$E_p$\end{rotate}}%
\psfrag{009}[tc][tc]{\matlabtextA $n$}%
%
%% </text>
%
%% <xtick>
%
\def\matlabfragNegXTick{\mathord{\makebox[0pt][r]{$-$}}}
\psfrag{000}[ct][ct]{\matlabtextA $1$}%
\psfrag{001}[ct][ct]{\matlabtextA $2$}%
\psfrag{002}[ct][ct]{\matlabtextA $3$}%
\psfrag{003}[ct][ct]{\matlabtextA $4$}%
%
%% </xtick>
%
%% <ytick>
%
\psfrag{004}[rc][rc]{\matlabtextA $0$}%
\psfrag{005}[rc][rc]{\matlabtextA $0.02$}%
\psfrag{006}[rc][rc]{\matlabtextA $0.04$}%
\psfrag{007}[rc][rc]{\matlabtextA $0.06$}%
%
%% </ytick>

%% file: Recovery_phi.tex
% Generated using matlabfrag
% Version: v0.6.16
% Version Date: 04-Apr-2010
% Author: Zebb Prime
%
%% <text>
%
\providecommand\matlabtextA{\color[rgb]{0.000,0.000,0.000}\fontsize{10}{10}\selectfont\strut}%
\psfrag{008}[cl][cl]{\matlabtextA SP: $\tilde{u}_r$}%
\psfrag{009}[cl][cl]{\matlabtextA SP: $\tilde{\boldsymbol{\omega}}_\perp$}%
\psfrag{010}[cl][cl]{\matlabtextA SP: $\tilde{\mathbf{u}}_\perp$ (linearized)}%
\psfrag{011}[cl][cl]{\matlabtextA SP: $\tilde{\mathbf{u}}_\perp$}%
\psfrag{012}[cl][cl]{\matlabtextA SVD}%
\psfrag{013}[bc][bc]{\matlabtextA \begin{rotate}{-90}$E_\phi$\end{rotate}}%
\psfrag{014}[tc][tc]{\matlabtextA $n$}%
%
%% </text>
%
%% <xtick>
%
\def\matlabfragNegXTick{\mathord{\makebox[0pt][r]{$-$}}}
\psfrag{000}[ct][ct]{\matlabtextA $1$}%
\psfrag{001}[ct][ct]{\matlabtextA $2$}%
\psfrag{002}[ct][ct]{\matlabtextA $3$}%
\psfrag{003}[ct][ct]{\matlabtextA $4$}%
%
%% </xtick>
%
%% <ytick>
%
\psfrag{004}[rc][rc]{\matlabtextA $0$}%
\psfrag{005}[rc][rc]{\matlabtextA $0.02$}%
\psfrag{006}[rc][rc]{\matlabtextA $0.04$}%
\psfrag{007}[rc][rc]{\matlabtextA $0.06$}%
%
%% </ytick>

%% file: Recovery_chi.tex
% Generated using matlabfrag
% Version: v0.6.16
% Version Date: 04-Apr-2010
% Author: Zebb Prime
%
%% <text>
%
\providecommand\matlabtextA{\color[rgb]{0.000,0.000,0.000}\fontsize{10}{10}\selectfont\strut}%
\psfrag{009}[cl][cl]{\matlabtextA SP: $\tilde{\omega}_r$}%
\psfrag{010}[cl][cl]{\matlabtextA SP: $\tilde{\boldsymbol{\omega}}_\perp$}%
\psfrag{011}[cl][cl]{\matlabtextA SP: $\tilde{\mathbf{u}}_\perp$ (linearized)}%
\psfrag{012}[cl][cl]{\matlabtextA SP: $\tilde{\mathbf{u}}_\perp$}%
\psfrag{013}[cl][cl]{\matlabtextA SVD}%
\psfrag{014}[bc][bc]{\matlabtextA \begin{rotate}{-90}$E_\chi$\end{rotate}}%
\psfrag{015}[tc][tc]{\matlabtextA $n$}%
%
%% </text>
%
%% <xtick>
%
\def\matlabfragNegXTick{\mathord{\makebox[0pt][r]{$-$}}}
\psfrag{000}[ct][ct]{\matlabtextA $1$}%
\psfrag{001}[ct][ct]{\matlabtextA $2$}%
\psfrag{002}[ct][ct]{\matlabtextA $3$}%
\psfrag{003}[ct][ct]{\matlabtextA $4$}%
%
%% </xtick>
%
%% <ytick>
%
\psfrag{004}[rc][rc]{\matlabtextA $0$}%
\psfrag{005}[rc][rc]{\matlabtextA $0.01$}%
\psfrag{006}[rc][rc]{\matlabtextA $0.02$}%
\psfrag{007}[rc][rc]{\matlabtextA $0.03$}%
\psfrag{008}[rc][rc]{\matlabtextA $0.04$}%
%
%% </ytick>

%% file: PeriodicArray_Nc.tex
% Generated using matlabfrag
% Version: v0.6.16
% Version Date: 04-Apr-2010
% Author: Zebb Prime
%
%% <text>
%
\providecommand\matlabtextA{\color[rgb]{0.000,0.000,0.000}\fontsize{10}{10}\selectfont\strut}%
\psfrag{015}[cl][cl]{\matlabtextA SP: $\tilde{u}_r$ \& $\tilde{\omega}_r$}%
\psfrag{016}[cl][cl]{\matlabtextA SP: $\tilde{\boldsymbol{\omega}}_\perp$}%
\psfrag{017}[cl][cl]{\matlabtextA SP: $\tilde{\mathbf{u}}_\perp$ (linearized)}%
\psfrag{018}[cl][cl]{\matlabtextA SP: $\tilde{\mathbf{u}}_\perp$}%
\psfrag{019}[cl][cl]{\matlabtextA SVD}%
\psfrag{020}[bc][bc]{\matlabtextA $|1-F_z/(1-\beta)L^3P|$}%
\psfrag{021}[tc][tc]{\matlabtextA $N_c$}%
%
%% </text>
%
%% <xtick>
%
\def\matlabfragNegXTick{\mathord{\makebox[0pt][r]{$-$}}}
\psfrag{000}[ct][ct]{\matlabtextA $1$}%
\psfrag{001}[ct][ct]{\matlabtextA $2$}%
\psfrag{002}[ct][ct]{\matlabtextA $3$}%
\psfrag{003}[ct][ct]{\matlabtextA $4$}%
\psfrag{004}[ct][ct]{\matlabtextA $5$}%
\psfrag{005}[ct][ct]{\matlabtextA $6$}%
\psfrag{006}[ct][ct]{\matlabtextA $7$}%
\psfrag{007}[ct][ct]{\matlabtextA $8$}%
%
%% </xtick>
%
%% <ytick>
%
\psfrag{008}[rc][rc]{\matlabtextA $0$}%
\psfrag{009}[rc][rc]{\matlabtextA $0.005$}%
\psfrag{010}[rc][rc]{\matlabtextA $0.01$}%
\psfrag{011}[rc][rc]{\matlabtextA $0.015$}%
\psfrag{012}[rc][rc]{\matlabtextA $0.02$}%
\psfrag{013}[rc][rc]{\matlabtextA $0.025$}%
\psfrag{014}[rc][rc]{\matlabtextA $0.03$}%
%
%% </ytick>

%% file: PeriodicArray_ra.tex
% Generated using matlabfrag
% Version: v0.6.16
% Version Date: 04-Apr-2010
% Author: Zebb Prime
%
%% <text>
%
\providecommand\matlabtextA{\color[rgb]{0.000,0.000,0.000}\fontsize{10}{10}\selectfont\strut}%
\psfrag{011}[cl][cl]{\matlabtextA SP: $\tilde{u}_r \;\&\; \tilde{\omega}_r$}%
\psfrag{012}[cl][cl]{\matlabtextA SP: $\tilde{\boldsymbol{\omega}}_\perp$}%
\psfrag{013}[cl][cl]{\matlabtextA SP: $\tilde{\mathbf{u}}_\perp$ (linearized)}%
\psfrag{014}[cl][cl]{\matlabtextA SP: $\tilde{\mathbf{u}}_\perp$}%
\psfrag{015}[bc][bc]{\matlabtextA $|1-F_z/(1-\beta)L^3P|$}%
\psfrag{016}[tc][tc]{\matlabtextA $r/a$}%
%
%% </text>
%
%% <xtick>
%
\def\matlabfragNegXTick{\mathord{\makebox[0pt][r]{$-$}}}
\psfrag{000}[ct][ct]{\matlabtextA $1$}%
\psfrag{001}[ct][ct]{\matlabtextA $1.2$}%
\psfrag{002}[ct][ct]{\matlabtextA $1.4$}%
\psfrag{003}[ct][ct]{\matlabtextA $1.6$}%
\psfrag{004}[ct][ct]{\matlabtextA $1.8$}%
\psfrag{005}[ct][ct]{\matlabtextA $2$}%
%
%% </xtick>
%
%% <ytick>
%
\psfrag{006}[rc][rc]{\matlabtextA $0$}%
\psfrag{007}[rc][rc]{\matlabtextA $0.025$}%
\psfrag{008}[rc][rc]{\matlabtextA $0.05$}%
\psfrag{009}[rc][rc]{\matlabtextA $0.075$}%
\psfrag{010}[rc][rc]{\matlabtextA $0.1$}%
%
%% </ytick>

%% file: PeriodicArray_raz.tex
% Generated using matlabfrag
% Version: v0.6.16
% Version Date: 04-Apr-2010
% Author: Zebb Prime
%
%% <text>
%
\providecommand\matlabtextA{\color[rgb]{0.000,0.000,0.000}\fontsize{10}{10}\selectfont\strut}%
\psfrag{011}[bc][bc]{\matlabtextA $|1-F_z/(1-\beta)L^3P|$}%
\psfrag{012}[tc][tc]{\matlabtextA $r/a$}%
%
%% </text>
%
%% <xtick>
%
\def\matlabfragNegXTick{\mathord{\makebox[0pt][r]{$-$}}}
\psfrag{000}[ct][ct]{\matlabtextA $1$}%
\psfrag{001}[ct][ct]{\matlabtextA $1.1$}%
\psfrag{002}[ct][ct]{\matlabtextA $1.2$}%
\psfrag{003}[ct][ct]{\matlabtextA $1.3$}%
\psfrag{004}[ct][ct]{\matlabtextA $1.4$}%
\psfrag{005}[ct][ct]{\matlabtextA $1.5$}%
%
%% </xtick>
%
%% <ytick>
%
\psfrag{006}[rc][rc]{\matlabtextA $0$}%
\psfrag{007}[rc][rc]{\matlabtextA $0.0025$}%
\psfrag{008}[rc][rc]{\matlabtextA $0.005$}%
\psfrag{009}[rc][rc]{\matlabtextA $0.0075$}%
\psfrag{010}[rc][rc]{\matlabtextA $0.01$}%
%
%% </ytick>

%% file: rot_couple.tex
% Generated using matlabfrag
% Version: v0.6.16
% Version Date: 04-Apr-2010
% Author: Zebb Prime
%
%% <text>
%
\providecommand\matlabtextA{\color[rgb]{0.000,0.000,0.000}\fontsize{10}{10}\selectfont\strut}%
\psfrag{015}[cl][cl]{\matlabtextA SP: $\tilde{u}_r \;\&\; \tilde{\omega}_r$}%
\psfrag{016}[cl][cl]{\matlabtextA SP: $\tilde{\boldsymbol{\omega}}_\perp$}%
\psfrag{017}[cl][cl]{\matlabtextA SP: $\tilde{\mathbf{u}}_\perp$ (linearized)}%
\psfrag{018}[cl][cl]{\matlabtextA SP: $\tilde{\mathbf{u}}_\perp$}%
\psfrag{019}[cl][cl]{\matlabtextA SVD}%
\psfrag{020}[bc][bc]{\matlabtextA $|1-L_{b}/L_{p}|$}%
\psfrag{021}[tc][tc]{\matlabtextA $N_c$}%
%
%% </text>
%
%% <xtick>
%
\def\matlabfragNegXTick{\mathord{\makebox[0pt][r]{$-$}}}
\psfrag{000}[ct][ct]{\matlabtextA $1$}%
\psfrag{001}[ct][ct]{\matlabtextA $2$}%
\psfrag{002}[ct][ct]{\matlabtextA $3$}%
\psfrag{003}[ct][ct]{\matlabtextA $4$}%
\psfrag{004}[ct][ct]{\matlabtextA $5$}%
\psfrag{005}[ct][ct]{\matlabtextA $6$}%
\psfrag{006}[ct][ct]{\matlabtextA $7$}%
\psfrag{007}[ct][ct]{\matlabtextA $8$}%
%
%% </xtick>
%
%% <ytick>
%
\psfrag{008}[rc][rc]{\matlabtextA $0$}%
\psfrag{009}[rc][rc]{\matlabtextA $0.02$}%
\psfrag{010}[rc][rc]{\matlabtextA $0.04$}%
\psfrag{011}[rc][rc]{\matlabtextA $0.06$}%
\psfrag{012}[rc][rc]{\matlabtextA $0.08$}%
\psfrag{013}[rc][rc]{\matlabtextA $0.1$}%
\psfrag{014}[rc][rc]{\matlabtextA $0.12$}%
%
%% </ytick>

%% file: rot_rOa.tex
% Generated using matlabfrag
% Version: v0.6.16
% Version Date: 04-Apr-2010
% Author: Zebb Prime
%
%% <text>
%
\providecommand\matlabtextA{\color[rgb]{0.000,0.000,0.000}\fontsize{10}{10}\selectfont\strut}%
\psfrag{012}[cl][cl]{\matlabtextA SP: $ \tilde{u}_r \;\&\; \tilde{\omega}_r$}%
\psfrag{013}[cl][cl]{\matlabtextA SP: $\tilde{\boldsymbol{\omega}}_\perp$}%
\psfrag{014}[cl][cl]{\matlabtextA SP: $\tilde{\mathbf{u}}_\perp$ (linearized)}%
\psfrag{015}[cl][cl]{\matlabtextA SP: $\tilde{\mathbf{u}}_\perp$}%
\psfrag{016}[bc][bc]{\matlabtextA $|1-L_{b}/L_{p}|$}%
\psfrag{017}[tc][tc]{\matlabtextA $r/a$}%
%
%% </text>
%
%% <xtick>
%
\def\matlabfragNegXTick{\mathord{\makebox[0pt][r]{$-$}}}
\psfrag{000}[ct][ct]{\matlabtextA $1$}%
\psfrag{001}[ct][ct]{\matlabtextA $1.2$}%
\psfrag{002}[ct][ct]{\matlabtextA $1.4$}%
\psfrag{003}[ct][ct]{\matlabtextA $1.6$}%
\psfrag{004}[ct][ct]{\matlabtextA $1.8$}%
\psfrag{005}[ct][ct]{\matlabtextA $2$}%
%
%% </xtick>
%
%% <ytick>
%
\psfrag{006}[rc][rc]{\matlabtextA $0$}%
\psfrag{007}[rc][rc]{\matlabtextA $0.05$}%
\psfrag{008}[rc][rc]{\matlabtextA $0.1$}%
\psfrag{009}[rc][rc]{\matlabtextA $0.15$}%
\psfrag{010}[rc][rc]{\matlabtextA $0.2$}%
\psfrag{011}[rc][rc]{\matlabtextA $0.25$}%
%
%% </ytick>

%% file: falling_re1.tex
% Generated using matlabfrag
% Version: v0.6.16
% Version Date: 04-Apr-2010
% Author: Zebb Prime
%
%% <text>
%
\providecommand\matlabtextA{\color[rgb]{0.000,0.000,0.000}\fontsize{10}{10}\selectfont\strut}%
\psfrag{011}[bc][bc]{\matlabtextA $a/h = 6, N_c = 2$}%
\psfrag{012}[bc][bc]{\matlabtextA $Re = 2aw/\nu$}%
\psfrag{013}[tc][tc]{\matlabtextA $t \nu/a^2$}%
%
%% </text>
%
%% <xtick>
%
\def\matlabfragNegXTick{\mathord{\makebox[0pt][r]{$-$}}}
\psfrag{000}[ct][ct]{\matlabtextA $0$}%
\psfrag{001}[ct][ct]{\matlabtextA $0.4$}%
\psfrag{002}[ct][ct]{\matlabtextA $0.8$}%
\psfrag{003}[ct][ct]{\matlabtextA $1.2$}%
\psfrag{004}[ct][ct]{\matlabtextA $1.6$}%
%
%% </xtick>
%
%% <ytick>
%
\psfrag{005}[rc][rc]{\matlabtextA $-25$}%
\psfrag{006}[rc][rc]{\matlabtextA $-20$}%
\psfrag{007}[rc][rc]{\matlabtextA $-15$}%
\psfrag{008}[rc][rc]{\matlabtextA $-10$}%
\psfrag{009}[rc][rc]{\matlabtextA $-5$}%
\psfrag{010}[rc][rc]{\matlabtextA $0$}%
%
%% </ytick>

%% file: falling_re2.tex
% Generated using matlabfrag
% Version: v0.6.16
% Version Date: 04-Apr-2010
% Author: Zebb Prime
%
%% <text>
%
\providecommand\matlabtextA{\color[rgb]{0.000,0.000,0.000}\fontsize{10}{10}\selectfont\strut}%
\psfrag{011}[cl][cl]{\matlabtextA SP: $\tilde{\boldsymbol{\omega}}_\perp$}%
\psfrag{012}[cl][cl]{\matlabtextA SP: $\tilde{\mathbf{u}}_\perp$ (linearized)}%
\psfrag{013}[cl][cl]{\matlabtextA SVD}%
\psfrag{014}[bc][bc]{\matlabtextA $a/h = 8, N_c = 3$}%
\psfrag{015}[tc][tc]{\matlabtextA $t \nu/a^2$}%
%
%% </text>
%
%% <xtick>
%
\def\matlabfragNegXTick{\mathord{\makebox[0pt][r]{$-$}}}
\psfrag{000}[ct][ct]{\matlabtextA $0$}%
\psfrag{001}[ct][ct]{\matlabtextA $0.4$}%
\psfrag{002}[ct][ct]{\matlabtextA $0.8$}%
\psfrag{003}[ct][ct]{\matlabtextA $1.2$}%
\psfrag{004}[ct][ct]{\matlabtextA $1.6$}%
%
%% </xtick>
%
%% <ytick>
%
\psfrag{005}[rc][rc]{\matlabtextA $-25$}%
\psfrag{006}[rc][rc]{\matlabtextA $-20$}%
\psfrag{007}[rc][rc]{\matlabtextA $-15$}%
\psfrag{008}[rc][rc]{\matlabtextA $-10$}%
\psfrag{009}[rc][rc]{\matlabtextA $-5$}%
\psfrag{010}[rc][rc]{\matlabtextA $0$}%
%
%% </ytick>

%% file: falling_t1.tex
% Generated using matlabfrag
% Version: v0.6.16
% Version Date: 04-Apr-2010
% Author: Zebb Prime
%
%% <text>
%
\providecommand\matlabtextA{\color[rgb]{0.000,0.000,0.000}\fontsize{10}{10}\selectfont\strut}%
\psfrag{011}[cl][cl]{\matlabtextA SP: $\tilde{\boldsymbol{\omega}}_\perp$}%
\psfrag{012}[cl][cl]{\matlabtextA SP: $\tilde{\mathbf{u}}_\perp$ (linearized)}%
\psfrag{013}[cl][cl]{\matlabtextA SVD}%
\psfrag{014}[bc][bc]{\matlabtextA Cost [sec]}%
\psfrag{015}[tc][tc]{\matlabtextA $t \nu/a^2$}%
\psfrag{016}[bc][bc]{\matlabtextA $a/h = 6, N_c = 2$}%
%
%% </text>
%
%% <xtick>
%
\def\matlabfragNegXTick{\mathord{\makebox[0pt][r]{$-$}}}
\psfrag{000}[ct][ct]{\matlabtextA $0$}%
\psfrag{001}[ct][ct]{\matlabtextA $0.4$}%
\psfrag{002}[ct][ct]{\matlabtextA $0.8$}%
\psfrag{003}[ct][ct]{\matlabtextA $1.2$}%
\psfrag{004}[ct][ct]{\matlabtextA $1.6$}%
%
%% </xtick>
%
%% <ytick>
%
\psfrag{005}[rc][rc]{\matlabtextA $0$}%
\psfrag{006}[rc][rc]{\matlabtextA $200$}%
\psfrag{007}[rc][rc]{\matlabtextA $400$}%
\psfrag{008}[rc][rc]{\matlabtextA $600$}%
\psfrag{009}[rc][rc]{\matlabtextA $800$}%
\psfrag{010}[rc][rc]{\matlabtextA $1000$}%
%
%% </ytick>

%% file: falling_t2.tex
% Generated using matlabfrag
% Version: v0.6.16
% Version Date: 04-Apr-2010
% Author: Zebb Prime
%
%% <text>
%
\providecommand\matlabtextA{\color[rgb]{0.000,0.000,0.000}\fontsize{10}{10}\selectfont\strut}%
\psfrag{011}[tc][tc]{\matlabtextA $t \nu/a^2$}%
\psfrag{012}[bc][bc]{\matlabtextA $a/h = 8, N_c = 3$}%
%
%% </text>
%
%% <xtick>
%
\def\matlabfragNegXTick{\mathord{\makebox[0pt][r]{$-$}}}
\psfrag{000}[ct][ct]{\matlabtextA $0$}%
\psfrag{001}[ct][ct]{\matlabtextA $0.4$}%
\psfrag{002}[ct][ct]{\matlabtextA $0.8$}%
\psfrag{003}[ct][ct]{\matlabtextA $1.2$}%
\psfrag{004}[ct][ct]{\matlabtextA $1.6$}%
%
%% </xtick>
%
%% <ytick>
%
\psfrag{005}[rc][rc]{\matlabtextA $0$}%
\psfrag{006}[rc][rc]{\matlabtextA $200$}%
\psfrag{007}[rc][rc]{\matlabtextA $400$}%
\psfrag{008}[rc][rc]{\matlabtextA $600$}%
\psfrag{009}[rc][rc]{\matlabtextA $800$}%
\psfrag{010}[rc][rc]{\matlabtextA $1000$}%
%
%% </ytick>

%% file: falling_force1.tex
% Generated using matlabfrag
% Version: v0.6.16
% Version Date: 04-Apr-2010
% Author: Zebb Prime
%
%% <text>
%
\providecommand\matlabtextA{\color[rgb]{0.000,0.000,0.000}\fontsize{10}{10}\selectfont\strut}%
\psfrag{010}[bc][bc]{\matlabtextA $(F_z-m_p g)/m_pg$}%
\psfrag{011}[bc][bc]{\matlabtextA $a/h = 6, N_c = 2$}%
\psfrag{012}[tc][tc]{\matlabtextA $t \nu/a^2$}%
%
%% </text>
%
%% <xtick>
%
\def\matlabfragNegXTick{\mathord{\makebox[0pt][r]{$-$}}}
\psfrag{000}[ct][ct]{\matlabtextA $0$}%
\psfrag{001}[ct][ct]{\matlabtextA $0.4$}%
\psfrag{002}[ct][ct]{\matlabtextA $0.8$}%
\psfrag{003}[ct][ct]{\matlabtextA $1.2$}%
\psfrag{004}[ct][ct]{\matlabtextA $1.6$}%
%
%% </xtick>
%
%% <ytick>
%
\psfrag{005}[rc][rc]{\matlabtextA $0.4$}%
\psfrag{006}[rc][rc]{\matlabtextA $0.6$}%
\psfrag{007}[rc][rc]{\matlabtextA $0.8$}%
\psfrag{008}[rc][rc]{\matlabtextA $1$}%
\psfrag{009}[rc][rc]{\matlabtextA $1.2$}%
%
%% </ytick>

%% file: falling_force2.tex
% Generated using matlabfrag
% Version: v0.6.16
% Version Date: 04-Apr-2010
% Author: Zebb Prime
%
%% <text>
%
\providecommand\matlabtextA{\color[rgb]{0.000,0.000,0.000}\fontsize{10}{10}\selectfont\strut}%
\psfrag{010}[cl][cl]{\matlabtextA SP: $\tilde{\boldsymbol{\omega}}_\perp$}%
\psfrag{011}[cl][cl]{\matlabtextA SP: $\tilde{\mathbf{u}}_\perp$ (linearized)}%
\psfrag{012}[cl][cl]{\matlabtextA SVD}%
\psfrag{013}[bc][bc]{\matlabtextA $a/h = 8, N_c = 3$}%
\psfrag{014}[tc][tc]{\matlabtextA $t \nu/a^2$}%
%
%% </text>
%
%% <xtick>
%
\def\matlabfragNegXTick{\mathord{\makebox[0pt][r]{$-$}}}
\psfrag{000}[ct][ct]{\matlabtextA $0$}%
\psfrag{001}[ct][ct]{\matlabtextA $0.4$}%
\psfrag{002}[ct][ct]{\matlabtextA $0.8$}%
\psfrag{003}[ct][ct]{\matlabtextA $1.2$}%
\psfrag{004}[ct][ct]{\matlabtextA $1.6$}%
%
%% </xtick>
%
%% <ytick>
%
\psfrag{005}[rc][rc]{\matlabtextA $0.4$}%
\psfrag{006}[rc][rc]{\matlabtextA $0.6$}%
\psfrag{007}[rc][rc]{\matlabtextA $0.8$}%
\psfrag{008}[rc][rc]{\matlabtextA $1$}%
\psfrag{009}[rc][rc]{\matlabtextA $1.2$}%
%
%% </ytick>

%% file: PeriodicArray_rcage.tex
% Generated using matlabfrag
% Version: v0.6.16
% Version Date: 04-Apr-2010
% Author: Zebb Prime
%
%% <text>
%
\providecommand\matlabtextA{\color[rgb]{0.000,0.000,0.000}\fontsize{10}{10}\selectfont\strut}%
\psfrag{012}[cl][cl]{\matlabtextA $R_c/a = 0.9$: $\tilde{\mathbf{u}}_\perp$ (linearized)}%
\psfrag{013}[cl][cl]{\matlabtextA $R_c/a = 0.9$: $\tilde{\boldsymbol{\omega}}_\perp$}%
\psfrag{014}[cl][cl]{\matlabtextA $R_c/a = 1$: $\tilde{\mathbf{u}}_\perp$ (linearized)}%
\psfrag{015}[cl][cl]{\matlabtextA $R_c/a = 1$: $\tilde{\boldsymbol{\omega}}_\perp$}%
\psfrag{016}[bc][bc]{\matlabtextA $|1-F_z/(1-\beta)L^3P|$}%
\psfrag{017}[tc][tc]{\matlabtextA $r/a$}%
%
%% </text>
%
%% <xtick>
%
\def\matlabfragNegXTick{\mathord{\makebox[0pt][r]{$-$}}}
\psfrag{000}[ct][ct]{\matlabtextA $1$}%
\psfrag{001}[ct][ct]{\matlabtextA $1.05$}%
\psfrag{002}[ct][ct]{\matlabtextA $1.1$}%
\psfrag{003}[ct][ct]{\matlabtextA $1.15$}%
%
%% </xtick>
%
%% <ytick>
%
\psfrag{004}[rc][rc]{\matlabtextA $0$}%
\psfrag{005}[rc][rc]{\matlabtextA $0.02$}%
\psfrag{006}[rc][rc]{\matlabtextA $0.04$}%
\psfrag{007}[rc][rc]{\matlabtextA $0.06$}%
\psfrag{008}[rc][rc]{\matlabtextA $0.08$}%
\psfrag{009}[rc][rc]{\matlabtextA $0.1$}%
\psfrag{010}[rc][rc]{\matlabtextA $0.12$}%
\psfrag{011}[rc][rc]{\matlabtextA $0.14$}%
%
%% </ytick>